\documentclass[twocolumn,showpacs,prb,amsfonts,amsmath,amssymb,floatfix,superscriptaddress]{revtex4}
\usepackage{color}
\usepackage{hhline}
\usepackage{mathrsfs}
\usepackage{graphicx}
\usepackage{dcolumn}
\usepackage{bm}
\usepackage{multirow,bigdelim}
\usepackage{booktabs}
\input{epsf}
\usepackage{amsmath}

\def\C60{{\rm C}_{60}}
\def\A3C60{{\rm A}_3{\rm C}_{60}}
\def\K3C60{{\rm K}_3{\rm C}_{60}}
\def\Rb3C60{{\rm Rb}_3{\rm C}_{60}}
\def\Cs3C60{{\rm Cs}_3{\rm C}_{60}}

\def\Tc{T_{\rm c}}

\def\Hscf1{{\cal H}_{\rm SCF}}
\def\dvscf1{\Delta V_{\rm SCF}}
\def\Hscfkq1{{\cal H}_{\rm SCF}^{{\mathbf k}+{\mathbf q}} }
\def\kq1{{\mathbf k} + {\mathbf q}}
\def\k1{{\mathbf k}}
\def\kp1{{\mathbf k}'}
\def\q1{{\mathbf q}}
\def\rp1{{\mathbf r}'}
\def\thetan1{\tilde{\theta}_{F,n}}
\def\thetam1{\tilde{\theta}_{F,m}}
\def\thetamn1{\tilde{\theta}_{m,n}}
\def\thetanm1{\tilde{\theta}_{n,m}}
\def\g1{{\mathbf G}}
\def\gp1{{\mathbf G}'}

\begin{document}

\title{{\em Ab initio} downfolding for electron-phonon coupled systems: constrained density-functional perturbation theory (cDFPT)}

\author{Yusuke Nomura}
\email{yusuke.nomura@riken.jp}
\affiliation{Department of Applied Physics, University of Tokyo, Hongo, Bunkyo-ku, Tokyo, 113-8656, Japan}
\altaffiliation[Present address: ]{Centre de Physique Th\'eorique, \'Ecole Polytechnique, CNRS, Universit\'e Paris-Saclay, F-91128 Palaiseau, France}

\author{Ryotaro Arita}
\affiliation{Center for Emergent Matter Science (CEMS), RIKEN, Hirosawa, Wako, Saitama 351-0198, Japan}
\affiliation{JST ERATO Isobe Degenerate $\pi$-Integration Project, AIMR, Tohoku University, 2-1-1 Katahira, Aoba-ku,
Sendai, 980-8577, Japan}

\date{\today}

\begin{abstract}
We formulate an {\it ab initio} downfolding scheme for electron-phonon coupled systems. 
In this scheme, we calculate partially renormalized phonon frequencies and electron-phonon coupling, which include the screening effects of high-energy electrons, to construct 
a realistic Hamiltonian consisting of low-energy electron and phonon degrees of freedom. 
We show that our scheme, which we call constrained density-functional perturbation theory (cDFPT), can be implemented by slightly modifying the conventional DFPT, which is one of the standard methods to calculate phonon properties from first principles. 
Our scheme can be applied to various phonon-related problems, such as superconductivity, electron and thermal transport, thermoelectricity, piezoelectricity, dielectricity and multiferroicity. 
We believe that the cDFPT provides a firm basis for the understanding of the role of phonons in strongly correlated materials.  
Here, we apply the scheme to the fullerene superconductors and discuss how the realistic low-energy Hamiltonian is constructed.
\end{abstract} 
\maketitle 


\section{Introduction}\label{Sec:intro}

A quantitative description of the strongly correlated materials is one of the most challenging goals in condensed matter physics. 
In particular, an accurate treatment for the lattice degrees of freedom in the strongly correlated regime is necessary for 
a description or even a prediction of functional materials such as high-transition-temperature (high-$\Tc$) superconductors, thermoelectrics, piezoelectrics, and multiferroics. 
However, the interplay between strong correlation and electron-phonon coupling has yet to be fully understood. 
For example, the role of the electron-phonon interaction in the cuprate superconductors is still controversial.~\cite{doi:10.1080/13642810208220725,Louie_cuprate_nature,0953-8984-20-4-043201,comment_nature,PhysRevB.82.064513}
Recently, it has been proposed that the 
electron correlation enhances the electron-phonon coupling.~\cite{PhysRevB.82.245409,PhysRevB.81.073106,PhysRevB.84.155104,PhysRevX.3.021011}
The phonon might cooperate with plasmons to realize high-$\Tc$ superconductivity.~\cite{PhysRevLett.111.057006,doi:10.7566/JPSJ.83.061016,PhysRevB.91.224513}
It has been shown that an unusual cooperation between multi-orbital electronic correlation and the Jahn-Teller phonons is the essence of high-$\Tc$ $s$-wave superconductivity next to the Mott insulating phase in the fullerides.~\cite{nomura_science_advances,Capone28062002,RevModPhys.81.943,PhysRevLett.90.167006} 

In this paper, we propose that a combination of the density functional theory (DFT) and model calculations, 
which is  one of the most powerful methods to study the strongly correlated materials,~\cite{RevModPhys.78.865,doi:10.1080/00018730701619647,doi:10.1143/JPSJ.79.112001}  
can also be powerful in studying the electron-phonon coupled systems with strong electron correlations.
This idea relies on the energy hierarchy in the electronic structure:~\cite{doi:10.1143/JPSJ.79.112001}
By the strong electronic correlation and the electron-phonon coupling, the low-energy bands near the Fermi level $E_F$, which we call target bands, 
may be heavily reconstructed,
while the structure of the high-energy bands will not change drastically. 
Furthermore, at a temperature where the low-energy phenomena (e.g., superconductivity) emerge,  
the high-energy states are nearly frozen, i.e., they are nearly totally occupied or empty.
Then, nearly all the excitation processes occur in $t$-subspace, the subspace which the target bands span
(for later use, we define $r$-subspace as the rest of the Hilbert space). 
The most important electron-phonon coupling processes are the couplings between these $t$-subspace electrons and phonons. 
Therefore, the low-energy physical properties are governed by the low-energy electrons and the phonons.  

This hierarchical structure allows us to construct the following three-stage scheme:~\cite{doi:10.1143/JPSJ.79.112001}
\begin{enumerate}
\item Obtain the global energy structure by the DFT and define the low-energy subspace.
\item Trace out the high-energy electron degrees of freedom and derive a low-energy effective Hamiltonian (downfolding).
         The degrees of freedom in the Hamiltonian consist of the $t$-subspace electrons and the phonons. 
\item Solve the derived model accurately by the model calculation method.
\end{enumerate}
In this scheme, we take into account the material dependence and the high-energy electronic structure by the DFT, 
and the effects of electron correlation and the electron-phonon coupling in the low-energy subspace ($t$-subspace) are considered by the model calculation. 
A key step in the scheme is the step 2, i.e., the dowonfolding procedure to derive the low-energy Hamiltonian.

When we restrict ourselves to the electron degrees of freedom and forget about the phonons, there has been much effort in the development of the downfolding scheme.
In this case, the low-energy Hamiltonian would consist of the electron one-body (hopping) and Coulomb interaction terms. 
By employing a localized basis such as the maximally localized Wannier function,~\cite{PhysRevB.56.12847,PhysRevB.65.035109,RevModPhys.84.1419} the derived model has a form of the extended Hubbard model. 
The one-body part describes a realistic hopping structure in the $t$-subspace. 
The effective interaction between the $t$-subspace electrons is a partially-screened Coulomb interaction.   
This is because the high-energy electrons, which are traced out, gives a renormalization of the Coulomb interaction. 
We refer to it as a ``partially"-screened interaction because it does not include screening processes originating from the $t$-subspace electrons, which are {\it not} traced out and remain as active degrees of freedom.
This partial screening is often calculated within the constrained random phase approximation (cRPA),~\cite{PhysRevB.70.195104}
which considers the screening effect of the high-energy electrons within the RPA. 

The downfolding scheme combined with the model-calculation method has been successfully applied to 
e.g., iron-based superconductors,~\cite{PhysRevB.83.144512,doi:10.1143/JPSJ.79.044705,PhysRevB.80.085101,yin_nphys,doi:10.1143/JPSJ.80.023704,misawa_ncom} cuprates,~\cite{PhysRevLett.105.057003,PhysRevB.80.054501,PhysRevB.82.125107,1367-2630-16-3-033009} transition metal oxides,~\cite{PhysRevB.72.155106,doi:10.1146/annurev-conmatphys-020911-125045} and organic compounds.~\cite{doi:10.1143/JPSJ.78.083710,doi:10.1143/JPSJ.81.034701}
Based on these successes, many attempts have been done to further improve the scheme. 
For example, there have been proposals to improve the one-body part,~\cite{PhysRevB.87.195144,PhysRevLett.113.266403} and the interaction part.~\cite{PhysRevB.86.085117,PhysRevB.82.045105,PhysRevB.92.045113,nphys_Udep,0295-5075-99-6-67003,PhysRevB.91.245156} 
Nowadays, GW-based (not DFT-based) scheme is also intensively studied.~\cite{PhysRevB.90.165138,0295-5075-100-6-67001,PhysRevB.88.235110,PhysRevB.88.165119}

Despite much effort in the electronic systems, the {\it ab initio} downfolding scheme for electron-phonon coupled systems has not been established. 
If we include the phonon degrees of freedom, the low-energy model acquires the electron-phonon coupling and phonon one-body terms in addition to the electron one-body and Coulomb interaction terms. 
As in the case of effective Coulomb interaction between the $t$-subspace, 
the electron-phonon coupling and phonon frequencies used in the low-energy Hamiltonian should be a partially renormalized quantity.~\cite{PhysRevB.84.184531}  
They are renormalized due to the coupling between the phonons and the high-energy electrons. 
The coupling between the phonons and $t$-subspace electrons is considered when we solve the model by the model-calculation method. 
When we derive the model, the renormalization originating from the $t$-subspace is excluded to avoid the double counting of it.
Basing on this idea, we recently proposed an {\it ab initio} scheme, called constrained density-functional perturbation theory (cDFPT).~\cite{PhysRevLett.112.027002} 

In this paper, we elaborate the practical detail of the cDFPT method. 
We show that the cDFPT method can be easily implemented by a slight modification of the conventional DFPT method, which is implemented in several {\it ab initio} packages. 
Then, we apply the scheme to the alkali-doped fullerides,~\cite{C60_super1,RevModPhys.69.575,fcc_science_advances}
where both the electron correlations and the electron-phonon interactions are important to explain the phase diagram.~\cite{nomura_science_advances} 
By comparing the cDFPT results with the DFPT results, we discuss how the partially screened quantities, which are used as an input for the model calculation, differ from the fully renormalized quantities.

This paper is organized as follows. 
In Sec.~\ref{sec:DFPT}, 
we review the DFPT~\cite{PhysRevB.43.7231,PhysRevB.51.6773,PhysRevB.60.11427,RevModPhys.73.515}
 to introduce our notation, since the cDFPT method is closely related to the conventional DFPT. 
Then, we move onto the the explanation of the cDFPT method in Sec.~\ref{meth:cDFPT}. 
There, we provide practical details to implement the cDFPT method. 
We also briefly compare the cDFPT method and another downfolding method proposed in Ref.~\onlinecite{PhysRevB.90.115435}. 
In Sec.~\ref{sec:result}, we show the cDFPT results for the alkali-doped fullerides and compare it with the DFPT results. 
Finally, in Sec.~\ref{sec:summary}, we give a summary of the paper. 



\section{review of density-functional perturbation theory}\label{sec:DFPT}

The cDFPT method is based on the DFPT method, where the fully renormalized electron-phonon coupling and 
phonon frequencies are calculated. 
As we will show below, the cDFPT method can be formulated as a slight modification of the DFPT algorithm. 
Here, we briefly review the DFPT method~\cite{PhysRevB.43.7231,PhysRevB.51.6773,PhysRevB.60.11427,RevModPhys.73.515}  just to introduce a notation used in the paper. 

\subsection{Phonon frequencies} \label{sec:DFPT_freq}

\subsubsection{Expression for interatomic force constants}
In solids, the phonon frequencies are determined by the following equation:~\cite{RevModPhys.73.515} 
\begin{eqnarray}
\label{Eq:omega_square}
\sum_{\kappa^{\prime} \alpha^{\prime}}
D_{\kappa \kappa^{\prime}}^{\alpha \alpha^{\prime}}({\bf q})
e_{\kappa^{\prime}}^{\alpha^{\prime}}({\bf q})
= \omega^{2}_{\bf q\nu}e_{\kappa}^{\alpha}({\bf q \nu})
\end{eqnarray}
with a momentum ${\bf q}$, the index for atoms $\kappa$, and the direction of the displacement $\alpha = \{ x, y, z \}$.  
This equation shows that the phonon frequency $\omega_{\bf q \nu}$ is given by the 
square root of the eigenvalues of the dynamical matrix $D({\bf q})$.
Since the dimension of the dynamical matrix  $D({\bf q})$ is $3n$ with $n$ being the number of the atoms in the unit cell, there exist $3n$ solutions (normal modes), which we label by the index $\nu$.
The eigenvectors of the dynamical matrix satisfy the orthonormality: 
\begin{eqnarray}
\sum_{\kappa \alpha}  e_{\kappa}^{\ast \alpha}({\bf q} \nu)
e_{\kappa}^{\alpha}({\bf q} \nu^{\prime}) = \delta_{\nu \nu^{\prime}}. 
\end{eqnarray}
The dynamical matrix is related to the interatomic force constants $C_{\kappa \kappa^{\prime}}^{\alpha \alpha^{\prime}}({\bf q})$ by 
\begin{eqnarray}
\label{Eq:Dmat}
D_{\kappa \kappa^{\prime}}^{\alpha \alpha^{\prime}}({\bf q}) = 
\frac{1}{\sqrt{\mathstrut M_{\kappa} M_{\kappa^{\prime}}}}
C_{\kappa \kappa^{\prime}}^{\alpha \alpha^{\prime}}({\bf q}), 
\end{eqnarray}
where $M_{\kappa}$ is the mass of the $\kappa$th atom. 
The interatomic force constants are written as~\cite{RevModPhys.73.515} 
\begin{eqnarray}\label{Eq:C}
 C_{\kappa \kappa'} ^ {\alpha \alpha'}({\bf q})  =  \frac{1}{N}
 \biggl [ && \! \! 
   \int \biggl(  \frac{\partial \rho (\bf r) }{ \partial u^{ \alpha}_{\kappa}  ({\bf q})} \biggr ) ^ {\ast}  
 \frac {\partial V_{\rm{ion}} (\bf r)} { \partial u^{  \alpha'}_{\kappa'} ({\bf q}) }  d{\bf r}   \nonumber  \\ 
&+&  \ \int \! \rho({\bf r})\frac{\partial^2 V_{\rm{ion}}({\bf r})} 
{\partial u^{\ast\alpha}_{\kappa} ({\bf q})\partial u^{\alpha'}_{\kappa'} ({\bf q})} d{\bf r} \nonumber \\
&+& \ \frac{\partial^2  E_{\rm N}}
{\partial u^{\ast\alpha}_{\kappa} ({\bf q})\partial u^{\alpha'}_{\kappa'} ({\bf q})}   \  \biggr ]_{u=0} 
\end{eqnarray}
with the number of the unit cells in the Born-von Karman boundary condition $N$, the displacement of the ion $u$,
the electron density $\rho$, the ionic potential $V_{\rm ion}$, and the Coulomb interaction energy among the nuclei $E_{\rm N}$. 
On the r.h.s. of Eq.~(\ref{Eq:C}), the first (second) term describes the contribution from the linear (quadratic) electron-phonon coupling and the third term describes the ionic contribution.\cite{note1}

\subsubsection{Electron density response}

In order to evaluate the interatomic force constants, we need to calculate the electron-density response  
to the ionic displacement $\partial \rho ({\bf r}) / \partial  u^{\alpha}_{\kappa}({\bf q})$, 
which is a key quantity in the cDFPT method, as we will show below. 
Before going into the explanation of the cDFPT, we show how the electron-density response is calculated in the usual DFPT method. 
Here, we consider a metallic case.~\cite{PhysRevB.51.6773}
In the DFT calculation for the metal, it is usual to introduce a smearing function $\tilde{\delta}(x)$ 
and the corresponding smoothed step function $\tilde{\theta}(x) = \int_{-\infty}^{x} \tilde{\delta}(x') dx'$.
In the present calculation, we employ the gaussian smearing $\tilde{\delta}(x) = {\rm exp}(-x^2)  \ \! / \! \sqrt{\pi}$.
Then, the expression for the electron density response  $\Delta \rho({\bf r})$  to the ionic displacement is given by  
\begin{eqnarray} \label{eq:deln_metal}
\Delta \rho ({\bf r})\  &=& \ \sum_{n,m} \frac{\thetan1 -\thetam1} { \varepsilon_n - \varepsilon_m} 
\psi_n^{\ast} ({\bf r}) \psi_m({\bf r}) 
\bigl \langle \psi_m \bigr | \dvscf1 \bigl | \psi_n \bigr \rangle \nonumber \\
  &=& \ 2 \sum_n \psi_n^{\ast} ({\bf r}) \Delta \psi_n({\bf r}), 
\end{eqnarray}
where we define $\Delta \psi_n({\bf r})$ as 
\begin{eqnarray} \label{Eq:delpsi_metal}
\Delta \psi_n ({\bf r}) = \sum_m 
\frac{ \thetan1 - \thetam1 } {\varepsilon_n - \varepsilon_m} \thetamn1 
  \psi_m ({\bf r}) 
\bigl \langle \psi_m \bigr | \dvscf1 \bigl | \psi_n \bigr \rangle. \nonumber \\ 
\end{eqnarray}
with composite indices for the band and the momentum $n,m$, the Kohn-Sham (KS) wave function $\psi_n$, and the KS eigenenergy $\varepsilon_n$. 
Here, $\thetan1$ and $\thetamn1$ are defined as  
$\thetan1 = \tilde{\theta}\bigl[  ( \varepsilon_F - \varepsilon_n ) / \sigma \bigr]$
and 
$\thetamn1 = \tilde{\theta}\bigl[  ( \varepsilon_m - \varepsilon_n ) / \sigma \bigr]$, respectively, 
with the Fermi energy $\varepsilon_F$ and a smearing width $\sigma$.
 In the actual calculation, the electron density response $\Delta \rho $ and the modulation of the potential $\Delta V_{\rm SCF}$ 
have indices of the momentum $\bf q$, the displaced atom $\kappa$, and the direction $\alpha$, which we omit for simplicity. 
 The change of the potential $\Delta V_{\rm SCF}$ due to the ionic displacement is given by a sum of the change of the ionic potential $\Delta V_{\rm ion}$ and the screening contribution from the Hartree and exchange channels (the second and third terms on the r.h.s. of the following equation): 
\begin{eqnarray}
\label{Eq:Vscf_ins}
\Delta V_{\rm SCF}({\bf r}) &=&  \Delta V_{\rm ion}({\bf r}) + 
e^{2} \int \frac{\Delta \rho({\bf r^{\prime}})}{|{\bf r}-{\bf r^{\prime}} | } d {\bf r^{\prime}}  \nonumber \\ 
&+&  \left. \frac{dV_{\rm xc}[\rho]}{d \rho} \right|_{\rho=\rho_{0}({\bf r)}} \Delta \rho({\bf r})
\end{eqnarray}
with $\rho_{0}$ being the electron density in the absence of the ionic displacement. 
Eqs.~(\ref{eq:deln_metal}) and (\ref{Eq:Vscf_ins}) are the equations to 
determine the electron density response, which are solved self-consistently. 

In the DFPT, in order to avoid the cumbersome summation over the unoccupied states in Eq.~(\ref{Eq:delpsi_metal}), 
one alternatively solves the following equations 
[Eqs. (72) and (73) in Ref.~\onlinecite{RevModPhys.73.515}]:
\begin{eqnarray} \label{eq:DFPT_metal}
\bigl( \Hscf1 + Q - \varepsilon_n \bigr) 
 \bigl | \Delta \psi_n \bigr \rangle
 = - \bigl (\thetan1 - P_n \bigr ) \dvscf1 \bigl | \psi_n \bigr \rangle \nonumber \\ 
\end{eqnarray}
where  
\begin{eqnarray}
Q = \sum_m  \alpha_m \bigl | \psi_m \bigr \rangle  \bigl \langle \psi_m \bigr |,  
\ P_n = \sum_m \beta_{n,m}    \bigl | \psi_m \bigr \rangle  \bigl \langle \psi_m \bigr |
\end{eqnarray}
with 
\begin{eqnarray}\label{eq.beta_normal}
\beta_{n,m} = \thetan1 \thetanm1 + \thetam1 \thetamn1 + \alpha_m \frac{ \thetan1 - \thetam1 } {\varepsilon_n - \epsilon_m} \thetamn1.  \nonumber \\ 
\end{eqnarray}
Here $\alpha_m$'s are parameters to avoid null eigenvalues of the $\bigl( \Hscf1 + Q - \varepsilon_n \bigr) $ matrix, which can be set to be a 
constant value which is larger than [(maximum energy among partial occupied states) $-$ (minimum energy of the occupied states)] for all the partially occupied states,
 and zero for the totally unoccupied states.~\cite{RevModPhys.73.515} 
This $\alpha_m$ parametrization enables the calculation without any information about the totally unoccupied states. 
In Appendix~\ref{app.omit_unocc}, we show that the solution of Eq.~(\ref{eq:DFPT_metal}) is indeed 
identical with that of Eq.~(\ref{Eq:delpsi_metal}). 

When the perturbation has the periodicity with the lattice ($\q1 = {\mathbf 0}$),
the Fermi energy may change and $\Delta \rho$ acquires an additional term:~\cite{RevModPhys.73.515} 
\begin{eqnarray}
\label{Eq.q_0_Ef1}
\Delta \rho ({\bf r}) = 2 \sum_n \psi_n^{\ast} ({\bf r}) \Delta \psi_n({\bf r}) + \rho({\bf r},\varepsilon_F) \Delta \varepsilon_F
\end{eqnarray}
with
\begin{eqnarray}
\label{Eq.q_0_Ef2}
\rho({\bf r}, \varepsilon) = \sum_n \frac{1}{\sigma} \tilde{\delta} \left(\frac{\varepsilon-\varepsilon_n}{\sigma}\right) \bigl |  \psi_n ({\bf r}) \bigr |^2.
\end{eqnarray}
The change in the Fermi energy $\Delta \varepsilon_F$ is determined by the charge neutrality condition.~\cite{RevModPhys.73.515}

\subsection{Electron-phonon coupling}

When the ions move from their equilibrium position, the ionic potential changes.  
Then, the surrounding electrons will respond to the potential change and screen it. 
The electron will feel this screened potential change and will be scattered. 
This process is expressed by the Hamiltonian 
\begin{eqnarray}
 \hat{{\mathcal H}}_{\rm el\mathchar`-ph}  =    \frac{1}{\sqrt{N}} 
\sum_{\q1\nu}  \! \sum_{\k1nn'\sigma} 
g^{ \nu}_{n' n}(\k1,\q1) c_{n'\kq1}^{\sigma \dagger } c_{n\k1}^{\sigma} 
( b_{\q1\nu} + b^{\dagger}_{-\q1\nu}). \nonumber \\ 
\end{eqnarray}
Here, 
\begin{eqnarray}
\label{Eq:g_elphfull}
g_{n^{\prime} n}^{\nu}({\bf k, q})  \ &=&   \ \sum_{\kappa\alpha}
 \sqrt{\frac{\hbar}{2 M_{\kappa} \omega_{{\bf q} \nu}} } \ 
   e^{\alpha}_{\kappa} ({\bf q}\nu )   \times  \nonumber \\ 
&&  \ \ \     \left\langle  \psi_{n' \kq1} \left|  \frac{\partial V_{\rm SCF} ({\bf r}) }{ \partial u^{\alpha}_{\kappa} ({\bf q}) }  \right| \psi_{n\k1} \right\rangle
\end{eqnarray}
  is the electron-phonon-coupling matrix element involving the Bloch states $\psi_{n{\bf k}}$ and $\psi_{n'{\bf k+q}}$ and the $\nu$th branch phonon with the wave vector $\bf q$.
$c_{n\k1}^{\sigma}$ ($c_{n\k1}^{\sigma\dagger}$) annihilates (creates) an electron on the $n$th Bloch orbital with the wave vector $\k1$ and the spin $\sigma$. 
$b_{\q1 \nu}$ ($b^{\dagger}_{\q1 \nu}$) is the annihilation (creation) operator for the phonon labeled by the $\nu$th branch and the momentum $\bf q$.  

\vspace{-0.3cm}
\section{Constrained density-functional perturbation theory}\label{meth:cDFPT}
\vspace{-0.2cm}
\subsection{Basic idea and practical implementation}
Our goal is to derive the low-energy Hamiltonian for the electron-phonon coupled systems,
which consists of the low-energy ($t$-subspace) electrons and the phonons. The Hamiltonian reads
\begin{eqnarray}
 \hat{{\mathcal H}}   =   \hat{{\mathcal H}}_{\rm el} + 
  \hat{{\mathcal H}}_{\rm el\mathchar`-el} + 
 \hat{{\mathcal H}}_{\rm el\mathchar`-ph} + 
  \hat{{\mathcal H}}_{\rm ph} +   \hat{{\mathcal H}}_{\rm DC},   
  \label{H_ele_ph_system}
\end{eqnarray}
where  $\hat{{\mathcal H}}_{\rm el}$ is the electronic one-body part (onsite energy and hopping terms), 
and $\hat{{\mathcal H}}_{\rm el\mathchar`-el}$ is the Coulomb interaction term, such as the Hubbard $U$.   
In this paper, we focus on the electron-phonon coupling $\hat{{\mathcal H}}_{\rm el\mathchar`-ph}$ and phonon one-body term $ \hat{{\mathcal H}}_{\rm ph}$, which are given by 
\begin{eqnarray}
\label{Eq:H_elph}
 \hat{{\mathcal H}}_{\rm el\mathchar`-ph} = 
\frac{1}{\sqrt{N}} 
\sum_{\q1\nu}  \! \sum_{\k1 ij \sigma} 
g^{(p) \nu}_{ij}(\k1,\q1) c_{i \kq1}^{\sigma \dagger } c_{j \k1}^{\sigma} 
( b_{\q1\nu} + b^{\dagger}_{-\q1\nu}). \nonumber \\ 
\end{eqnarray}
and 
\begin{eqnarray}
\label{Eq:H_ph}
\hat{{\mathcal H}}_{\rm ph} = \sum_{\q1 \nu}
\omega^{(p)}_{\q1 \nu}    b^{\dagger}_{\q1\nu}  b_{\q1\nu},  
\end{eqnarray}
respectively.  Here, we employ the Wannier gauge for the electronic degrees freedom labeled by $i, j$, 
since it is convenient for the low-energy solvers to take the Wannier gauge.
$\hat{{\mathcal H}}_{\rm DC}$ is a double counting correction, which is discussed in detail in Sec.~\ref{sec:DC}.
In this section, we show how the phonon frequencies $\omega^{(p)}$ and the electron-phonon coupling $g^{(p)}$ in the low-energy model should be parametrized.~\cite{PhysRevLett.112.027002}
As in the case of the effective Coulomb interactions 
in $\hat{{\mathcal H}}_{\rm el\mathchar`-el}$ calculated by the cRPA method,~\cite{PhysRevB.70.195104} they should be partially renormalized quantities, which take into account the  
renormalization effects associated with the elimination of the high-energy degrees of freedom(see Appendix~\ref{App:comp} for the comparison between the cDFPT and the cRPA).
In other words, we derive the parameters
with avoiding the double counting of the renormalization effects which are to be taken into account in the model analysis step.
To make it clear that these are partially renormalized quantities, we attach the superscript $(p)$. 

In the following, we discuss how the partially renormalized phonon quantities are calculated from first principles. 
For the partial renormalization, we first define the bare phonon frequencies and electron-phonon coupling. We then divide the renormalization processes into the low-energy contribution, which is to be excluded to realize the partial renormalization, and the rest of the contribution, which involves the high-energy electrons.

First, we consider the phonon frequencies. As we see in Sec.~\ref{sec:DFPT_freq}, the interatomic force constants [Eq.~(\ref{Eq:C})], which give the phonon frequencies, 
consist of several contributions.  
Since the low-energy Hamiltonian in Eq.~(\ref{H_ele_ph_system}) has the linear electron-phonon coupling term, which gives a renormalization of the phonon frequencies,
we define (ionic contribution) + (contribution from the quadratic electron-phonon coupling) as ``bare'' term, and (contribution from the linear electron-phonon coupling) as ``renormalizing" term.
Then the interatomic force constants $C_{\kappa \kappa'} ^ {\alpha \alpha'}({\bf q})$ given in Eq.~(\ref{Eq:C}) can be divided as $C_{\kappa \kappa'} ^ {\alpha \alpha'}({\bf q}) = \phantom{}^{\rm{bare}}C_{\kappa \kappa'} ^ {\alpha \alpha'}({\bf q}) + \phantom{}^{\rm{ren.}}C_{\kappa \kappa'} ^ {\alpha \alpha'}({\bf q})$, where 
$\phantom{}^{\rm{bare}}C_{\kappa \kappa'} ^ {\alpha \alpha'}({\bf q})$ gives the ``bare" phonon frequencies 
\begin{eqnarray}
\label{Eq.bare_freq}
\phantom{}^{\rm{bare}}C^{\alpha\alpha'}_{\kappa\kappa'}({\bf q})  =  \frac{1}{N}  \biggl [  &&
\frac{\partial^2 E_{\rm{N}}  } 
{\partial u^{\ast\alpha}_{\kappa} ({\bf q})\partial u^{\alpha'}_{\kappa'} ({\bf q})} \nonumber \\ 
 &+&   \int    \rho({\bf r})\frac{\partial^2 V_{\rm{ion}}({\bf r})}  
{\partial u^{\ast\alpha}_{\kappa} ({\bf q})\partial u^{\alpha'}_{\kappa'} ({\bf q})} d{\bf r} \biggr ],
\end{eqnarray}
and  $\phantom{}^{\rm{ren.}}C_{\kappa \kappa'} ^ {\alpha \alpha'}({\bf q})$ gives the renormalization of the phonon frequencies through the linear electron-phonon coupling
\begin{eqnarray}
\label{Eq:SE_ph}
\phantom{}^{\rm{ren.}} C_{\kappa \kappa'} ^ {\alpha \alpha'}({\bf q}) =  \frac{1}{N}
  \int \biggl(  \frac{\partial \rho (\bf r) }{ \partial u^{ \alpha}_{\kappa}  ({\bf q})} \biggr ) ^ {\ast} 
 \frac {\partial V_{\rm{ion}} (\bf r)} { \partial u^{  \alpha'}_{\kappa'} ({\bf q}) }  d{\bf r}.
\end{eqnarray}

Next, we consider the bare and renormalizing contributions to electron-phonon coupling [Eq.~(\ref{Eq:g_elphfull})]. 
The derivative of the self-consistent field potential $\partial V_{\rm{SCF}} ({\bf r}) / \partial  u^{\alpha}_{\kappa}({\bf q})$ in Eq.~(\ref{Eq:g_elphfull}) is also decomposed into the bare contribution
\begin{eqnarray} 
\label{Eq.bare_Vscf}
{\phantom{\Bigr]}}^{\rm{bare}} \biggl[  \frac{\partial V_{\rm{SCF}} ({\bf r}) }{\partial  u^{\alpha}_{\kappa} ({\bf q}) } \biggr] = 
    \frac{\partial V_{\rm{ion}} ({\bf r}) }{\partial  u^{\alpha}_{\kappa} ({\bf q}) }  
\end{eqnarray}
and the screening contribution (the change of the Hartree and exchange potentials)
\begin{eqnarray}
\label{Eq:delV}
{\phantom{\Bigr]}}^{\rm{ren.}} \biggl[  \frac{\partial V_{\rm{SCF}} ({\bf r}) }{\partial  u^{\alpha}_{\kappa} ({\bf q}) } \biggr] = 
\int &\biggl(& \frac { e^2 } {|{\bf r}-{\bf r^{\prime}} | }    +  
  \frac{dV_{\rm{xc}}({\bf r})}{d\rho}  \delta ({\bf r - \bf r'})     \biggr ) \nonumber \\
 &\times&   \ \frac{\partial \rho ({\bf r'}) }{\partial u^{\alpha}_{\kappa}({\bf q}) }
\ \!  d {{\bf r}'}. 
\end{eqnarray}

We see that the origin of the renormalization of the phonon frequencies and the screening for the electron-phonon couplings is the coupling between the lattice and the electrons, and the resulting modulation of the electron density due to the lattice displacement $\partial \rho ({\bf r}) / \partial  u^{\alpha}_{\kappa}({\bf q})$.  
The electron-density modulation $\partial \rho ({\bf r}) / \partial  u^{\alpha}_{\kappa}({\bf q})$ calculated in the conventional DFPT scheme is a sum of the contributions from all the possible particle-hole excitations [Eq.~(\ref{eq:deln_metal})].
In the cDFPT method,~\cite{PhysRevLett.112.027002} we exclude the target$\leftrightarrow$target excitation processes from the sum
in the calculation of the electron-density modulation. 
We use the resulting electron-density modulation for the renormalization contributions in Eqs.~(\ref{Eq:SE_ph}) and (\ref{Eq:delV}), 
which are added to the bare contributions in Eqs.~(\ref{Eq.bare_freq}) and (\ref{Eq.bare_Vscf}). 
This procedure gives the partially renormalized phonon frequencies and the electron-phonon couplings. 

Now, we propose a practical way to exclude the  target$\leftrightarrow$target processes from Eqs.~(\ref{eq:deln_metal}) and (\ref{eq:DFPT_metal}), the equations which determine the change of the electron density. 
If $\bigl | \psi_n \bigr \rangle$ in Eq.~(\ref{eq:DFPT_metal}) belongs to the $t$-subspace, 
in order to exclude the target$\leftrightarrow$target polarization processes, 
the r.h.s. of Eq.~(\ref{eq:DFPT_metal}) should be modified as 
\begin{eqnarray} \label{eq:DFPT_c}
\bigl( \Hscf1 + Q - \varepsilon_n \bigr) 
 \bigl | \Delta \psi_n \bigr \rangle
 =  - P_r \bigl (\thetan1 - P_n \bigr ) \dvscf1 \bigl | \psi_n \bigr \rangle \nonumber \\ 
\end{eqnarray}
with $P_r$ being the projection onto the $r$-subspace.
The very same constraint can be achieved by solving Eq.~(\ref{eq:DFPT_metal})
with modified $\beta_{n,m}$'s ($\tilde{\beta}_{n,m}$'s) given by  
\begin{widetext}
\begin{eqnarray}
\label{Eq.betanm}
\tilde{\beta}_{n,m} = 
\begin{cases} 
\; \ \ \thetan1     \hspace{0.6cm}  \bigl ( n,m \in t\rm{\mathchar`-subspace} \bigr ), \\ 
\; \ \  \thetan1 \thetanm1 + \thetam1 \thetamn1 + \alpha_m \frac{ \thetan1 - \thetam1 } {\varepsilon_n - \varepsilon_m} \thetamn1     \hspace{0.5cm} \bigl (\text{the other cases}\bigr ).
\end {cases} 
\end{eqnarray}
\end{widetext}
Note that in the latter case, $\tilde{\beta}_{n,m}$ has 
exactly the same form 
as that of Eq.~(\ref{eq.beta_normal}), i.e., $\tilde{\beta}_{n,m} = \beta_{n,m}$. 
Only when  $ n,m \in t\rm{\mathchar`-subspace}$, $\beta_{n,m}$ is modified. 
We can easily show that the r.h.s. of Eq.~(\ref{eq:DFPT_c}) with the original $\beta_{n,m}$'s
is equal to that of Eq.~(\ref{eq:DFPT_metal}) with $\tilde{\beta}_{n,m}$'s, 
which ensures the equivalence of the two types of the modifications. 
Using $\tilde{\beta}_{n,m}$ in Eq.~(\ref{Eq.betanm}) is also useful to exclude the contribution to the electron-density modulation from the possible change in the Fermi energy in the case of ${\bf q}={\bf 0}$ [the additional contribution given in Eqs.~(\ref{Eq.q_0_Ef1}) and (\ref{Eq.q_0_Ef2})]. 
The possible change in the Fermi energy originates from the intraband transitions at the Fermi level, which are the transition processes in the $t$-subspace and hence are excluded by employing $\tilde{\beta}_{n,m}$. 

When we consider the practical implementation, if one has a code of the conventional DFPT, 
it is easier to modify $\beta_{n,m}$ into $\tilde{\beta}_{n,m}$ than to employ Eq.~(\ref{eq:DFPT_c}).
One has only to modify the part where the $\beta_{n,m}$ parameters are defined, and no modification is needed in the other parts. 
In Appendix~\ref{app:implement}, we propose an example how we 
modify a source code to introduce 
$\tilde{\beta}_{n,m}$ 
in the case of {\sc quantum espresso} package.~\cite{0953-8984-21-39-395502,QEspresso}
With $\tilde{\beta}_{n,m}$'s and following the very same flow of calculations of the usual DFPT method, one can calculate the electron density response to the ionic displacement without target$\leftrightarrow$target polarization processes.
Then, with the resulting electron density response, we evaluate the partially-renormalized quantities $\omega^{(p)}$ and $g^{(p)}$. 

\subsection{Relation between fully and partially renormalized quantities}\label{sec_relate_p_f}

In this section, we show the relation between the partially and fully renormalized quantities.
The partially (fully) renormalized quantities are calculated by the cDFPT (conventional DFPT) method. 
The electron density response $\Delta \rho$ to the change of the ionic potential $\Delta V_{\rm{ion}}$ (bare perturbation) is given by~\cite{note2}
\begin{eqnarray} 
\label{Eq:deln}
\Delta \rho &=&  \underbrace{ \chi^0  \left(1-\tilde{v}\chi^0\right)^{-1}}_{\text{\normalsize$=\chi_{\rm{DFT}}$}}  \Delta V_{\rm{ion}}  \\  
\label{Eq:s2-KN}
&=& \chi^0 \Delta V_{\rm{SCF}},  
\end{eqnarray} 
where $\Delta V_{\rm{SCF}}$ is the screened potential change, given by 
\begin{eqnarray}
\label{Eq:s3-KN}
\Delta V_{\rm{SCF}}=\left(1-\tilde{v}\chi^0 \right)^{-1}\Delta V_{\rm{ion}}. 
\end{eqnarray}
Here, $\tilde{v}$ is given by $\tilde{v}= v+K_{\rm{xc}}$ with the bare Coulomb interaction $v$ and the exchange-correlation kernel $K_{\rm{xc}}=\delta V_{\rm{xc}}/\delta \rho$ ($V_{\rm{xc}}$ is the exchange-correlation potential).
Note that Eqs.~(\ref{Eq:s2-KN}) and (\ref{Eq:s3-KN}) correspond to Eqs.~(\ref{eq:deln_metal}) and (\ref{Eq:Vscf_ins}), respectively.
The screening expressed in Eq.~(\ref{Eq:s3-KN}) can be divided into two screening steps: 
One involving the high-energy degrees of freedom 
\begin{eqnarray}
\label{Eq:Vscfp}
\Delta V_{\rm{SCF}}^{(p)} =\left(1-\tilde{v}\chi^0_r\right)^{-1} \Delta V_{\rm{ion}}    
\end{eqnarray}
and the other associated with the target-target processes 
\begin{eqnarray}
\label{Eq:Vscff}
\Delta V_{\rm{SCF}}^{(f)}
=\left(1-\tilde{W}^{(p)}\chi^0_t\right)^{-1}  \Delta V_{\rm{SCF}}^{(p)} .        
\end{eqnarray}
Here, the total irreducible polarization $\chi^0$ is divided into $\chi^0_t$ and $\chi^0_r$ with the polarization within the $t$-subspace 
$\chi^0_t$ and the rest of the polarization $\chi^0_r = \chi^0 - \chi^0_t$. 
We have introduced the superscript $p$, and $f$ to explicitly distinguish between the partially ($p$) and fully ($f$) renormalized quantities. 
 $\tilde{W}^{(p)}$ is the partially screened Coulomb interaction given by 
\begin{eqnarray}
\tilde{W}^{(p)}=\left(1-\tilde{v}\chi^0_r\right)^{-1}\tilde{v}.
\label{Wp}
\end{eqnarray}
Since the electron-phonon coupling $g$ represents the scattering of the electrons by $\Delta V_{\rm{SCF}}$,
the screening process for the electron-phonon coupling can be decomposed in the very same way as that of $\Delta V_{\rm{SCF}}$ [Eqs.~(\ref{Eq:Vscfp}) and (\ref{Eq:Vscff})]; that is, $g^{(f)}=\left(1-\tilde{v}\chi^0\right)^{-1}g^{(b)}$ is decomposed into 
\begin{eqnarray}
\label{Eq:gp}
g^{(p)}=\left(1-\tilde{v}\chi^0_r\right)^{-1}g^{(b)}       
\end{eqnarray}
and 
\begin{eqnarray}
\label{Eq:gf}
g^{(f)}=\left(1-\tilde{W}^{(p)}\chi^0_t\right)^{-1}g^{(p)}.        
\end{eqnarray}
Eq.~(\ref{Eq:gf}) tells us that 
when we take into account the target-target screening processes at the DFT level for the model with the partially-screened Coulomb and electron-phonon interactions, we come back to the fully-screened electron-phonon interactions. 

The similar decomposition also applies to the renormalization of the phonon frequencies.
 In this case, the phonon self-energy is decomposed. 
 The renormalizing contribution to the interatomic force constants in Eq.~(\ref{Eq:SE_ph}) can be recast as
\begin{eqnarray}
^{\rm{ren.}}C = | g'^{(b)} | ^2 \chi_{\rm{DFT}}, 
\end{eqnarray}
where $g'^{(b)} = \sqrt{2M \omega^{(b)}}g^{(b)}$ with $\omega^{(b)}$ being the bare phonon frequency.
For simplicity, we have omitted the indices and represent the masses of the nucleus by a single mass $M$. 
We define the phonon self-energy in the DFPT scheme as 
\begin{eqnarray}
\label{Eq.sigma}
\Sigma = \frac{\phantom{}^{\rm{ren.}}C}{2M \omega^{(b)} }= | g^{(b)} | ^2 \chi_{\rm{DFT}}
\end{eqnarray}
The contribution to the phonon self-energy can be divided into $\Sigma_t$ and $\Sigma_r$, i.e., 
\begin{eqnarray}
\Sigma = \Sigma_t + \Sigma_r. 
\end{eqnarray}
Here, 
$\Sigma_r = | g^{(b)} | ^2 \chi^{r}_{\rm{DFT}}$
with $\chi^{r}_{\rm{DFT}} = \chi^0_{r}  \bigl(1-\tilde{v}\chi^0_{r} \bigr)^{-1}$  denotes the phonon self-energy due to the 
electron-phonon coupling involving the $r$-subspace electrons. 
The other part of the self-energy $\Sigma_t = | g^{(p)} | ^2 \chi^{t}_{\rm{DFT}}$ with $\chi^{t}_{\rm{DFT}} = \chi^0_{t}  \bigl(1-\tilde{W}^{(p)}\chi^0_{t} \bigr)^{-1}$ originates from the coupling between the $t$-subspace electrons and the phonons through the partially-screened coupling $g^{(p)}$.
See Appendix \ref{app:proof_S} for the proof that $\Sigma_t + \Sigma_r =  | g^{(p)} | ^2 \chi^{t}_{\rm{DFT}} + | g^{(b)} | ^2 \chi^{r}_{\rm{DFT}}$ is indeed identical to $\Sigma = | g^{(b)} | ^2 \chi_{\rm{DFT}}$.
The decomposition of $\Sigma$ into $\Sigma_t$ and $\Sigma_r$ corresponds to the division of the density-response contribution to $\phantom{}^{\rm{ren.}}C$ into the target-target contribution and the others, as the cDFPT scheme does.
With the decomposition of $\Sigma$, we can define the partially-dressed phonon Green's function $D^{(p)}$ as 
\begin{eqnarray}
 [ D^{(p)} ]^{-1} = [D^{(b)}] ^{-1} - \Sigma_{r}, 
\end{eqnarray}
with the bare phonon Green's function $D^{(b)}$. The bare phonon frequency $\omega^{(b)}$ is given by the pole of $D^{(b)}$. Similarly, the phonon frequency $\omega^{(p)}$ in the low-energy Hamiltonian is given by the pole of $D^{(p)}$.
If we further consider $\Sigma_t$, we obtain the fully-dressed phonon Green's function $D^{(f)}$ as 
\begin{eqnarray}
 [ D^{(f)} ]^{-1} = [D^{(p)}] ^{-1} - \Sigma_{t}. 
\end{eqnarray}

\subsection{Flow of the calculation and practical issues}
As we already mentioned, the flow to the cDFPT calculation just follows that of the usual DFPT. 
The difference comes from the setting of $\beta_{n,m}$ parameters. 
The flow of the calculation is as follows: 
\begin{enumerate}
\item Optimize the atomic positions within the DFT. 
\item Calculate the global energy structure by the DFT for the optimized structure and choose the target subspace for which we construct an effective Hamiltonian. 
\item Set $\tilde{\beta}_{n,m}$ parameters according to Eq.~(\ref{Eq.betanm}). 
\item  Perform the phonon calculation with $\tilde{\beta}_{n,m}$ parameters (the procedure is the very same as 
the conventional DFPT case).  \\ 
  $\rightarrow$ Obtain the partially renormalized phonon frequencies $\omega^{(p)}_{\q1 \nu}$  to be used in Eq.~(\ref{Eq:H_ph})
     and the partially screened potential change  $ \frac{\partial V^{(p)}_{\rm{SCF}} ({\bf r}) }{\partial  u^{\alpha}_{\kappa} ({\bf q}) }$. 
 \item Take the Wannier matrix element of  $ \frac{\partial V^{(p)}_{\rm{SCF}} ({\bf r}) }{\partial  u^{\alpha}_{\kappa} ({\bf q}) }$ to obtain the partially renormalized electron-phonon coupling term in Eq.~(\ref{Eq:H_elph}) as follows: 
 \begin{eqnarray}
\label{Eq:g_elphpatiall}
g_{ij }^{(p) \nu}({\bf k, q})  \ &=&   \ \sum_{\kappa\alpha}
 \sqrt{\frac{\hbar}{2 M_{\kappa} \omega^{(p)}_{{\bf q} \nu}} } \ 
   e^{(p)\alpha}_{\kappa} ({\bf q}\nu )   \times  \nonumber \\ 
&&  \ \ \     \left\langle  \psi^{(w)}_{i \kq1} \left|  \frac{\partial V^{(p)}_{\rm SCF} ({\bf r}) }{ \partial u^{\alpha}_{\kappa} ({\bf q}) }  \right| \psi^{(w)}_{j\k1} \right\rangle, 
\end{eqnarray}
where we use the superscript $(w)$ to make it clear that the wavefunction is in the Wannier gauge. 
\end{enumerate}

Finally, we mention one practical issue in obtaining the partially screened phonon frequencies $\omega^{(p)}_{\q1 \nu}$. 
In obtaining the fully renormalized phonon frequencies, we often impose the acoustic sum rule to 
ensure that the frequency of the acoustic phonon at $\q1 = {\bf 0}$ is zero. 
To obtain the partially renormalized phonon frequencies, we impose the same correction of the acoustic sum rule as that used in the calculation of the fully renormalized phonon frequencies.  
Then, the partially renormalized phonon frequency of the acoustic phonon at $\q1 = {\bf 0}$ does not always go to zero. 
This is because the phonon self-energy involving $t$-subspace electrons  $\Sigma_t = | g^{(p)} | ^2 \chi^{t}_{\rm{DFT}}$ can be finite, since there can be a finite coupling between the acoustic phonon and $t$-subspace electrons through the umklapp (${\bf G} \neq {\bf 0}$) processes, while the coupling for $\q1 + {\bf G} = {\bf 0}$ process is zero. 
We also give another explanation for possible non-zero phonon frequency for the acoustic mode at $\q1 = {\bf 0}$ . 
For example, in the case where the unit cell consists of a single atom, at ${\bf q} = {\bf 0}$, 
the ionic contribution to the interatomic force constant [the third term on the r.h.s. of Eq.~(\ref{Eq:C})] is zero. 
The first (second) term on the r.h.s. of Eq.~(\ref{Eq:C}), which is related with the linear (quadratic) electron-phonon coupling, gives negative (positive) contribution to the interatomic force constant. 
Since the first and second terms cancel with each other, the fully-renormalized phonon frequency at $\q1 = {\bf 0}$ goes to zero. 
In the cDFPT, we exclude the target contribution to the first term, thus imbalance occurs between the first and second terms, which makes the partially-renormalized phonon frequency nonzero.

\subsection{Double counting correction}\label{sec:DC}
When we combine the DFT and the model-calculation methods, we usually need a double counting correction.  
In the case of our scheme, we have a double counting problem for a possible change of the equilibrium positions of the atoms
due to the coupling between the lattice and the $t$-subspace electrons. 
The low-energy Hamiltonian should be formulated such that we obtain the equilibrium positions of the ions which agree with the optimized positions within the DFT level, after we solve the model at the static mean-field (DFT) level. 
To realize this, we need a double counting correction in the low-energy Hamiltonian, whose form is 
\begin{eqnarray}
\label{Eq.DC}
 \hat{{\mathcal H}}_{\rm DC}  =  -   \frac{1}{\sqrt{N}} 
\sum_{\nu}  \! \sum_{\k1ij \sigma} 
g^{ (p) \nu}_{ij}(\k1,\q1 \! =\! {\bf 0})  \langle c_{i \k1}^{\sigma \dagger } c_{j \k1}^{\sigma}  \rangle
( b_{{\bf 0}\nu} + b^{\dagger}_{{\bf 0} \nu}). \nonumber \\ 
\end{eqnarray}
Here, $ \langle c_{i \k1}^{\sigma \dagger } c_{j \k1}^{\sigma}  \rangle $ is the expectation value evaluated within the DFT. 

To understand the physical meaning of the double counting correction, we consider a simple case, where the $t$-subspace consists of a single band and only one Holstein phonon couples to the electron locally.  
Then, the electron-phonon coupling term in Eq.~(\ref{Eq:H_elph}) is given by 
\begin{eqnarray}
\label{Eq:H_elphsimgle}
 \hat{{\mathcal H}}_{\rm el\mathchar`-ph} =     \sum_{l } g^{(p)} n_{l } x_{l },  
\end{eqnarray}
where we switch to the real space representation and $l$ is the site index.
$n_l$ is the density operator for the site $l$ and $x_l$ is the displacement of the lattice.  
The double counting correction [Eq.~(\ref{Eq.DC})] becomes 
\begin{eqnarray}
\label{Eq:DC_simple}
\hat{{\mathcal H}}_{\rm DC}  =   - \sum_{l }  g^{(p)} \langle n_{l } \rangle  x_l.
\end{eqnarray}
If we 
put together the electron-phonon coupling, double counting, and potential energy terms
[the phonon-related part of the low-energy Hamiltonian in Eq.~(\ref{H_ele_ph_system})], it is given by 
\begin{eqnarray}
\label{Eq:simple_phonon}
   \hat{{\mathcal H}}  &=&  \sum_{l } g^{(p)} (n_{l } - \langle n_{l } \rangle   ) x_{l } + \sum_l  \frac{1}{2} ( \omega^{(p)})^2 x_l ^2 \nonumber \\
    &=&  \sum_{l } g^{(p)} n_{l }   x_{l } \! + \! \sum_l  \frac{1}{2} ( \omega^{(p)})^2 ( x_l - x^0_l )^2  + {\rm const.}
\end{eqnarray} 
where $x^0_l =  g^{(p)} \langle n_l \rangle  / ( \omega^{(p)})^2 $.  
In the above expression, we take the atomic mass to be 1 for simplicity. 
Now, the physical meaning of the double counting correction becomes clear: 
It gives the shift of the potential minimum of the lattice vibration by $x^0_l$, which is proportional to the occupation of the electron $ \langle n_l \rangle$. 
$x^0_l$ gives the equilibrium position of the lattice vibration without the effect of the low-energy electron manifold. 
When we solve the model at the mean-field (DFT) level, the equilibrium position goes back to the optimized position within the DFT, 
since the contributions from Eqs. (\ref{Eq:H_elphsimgle}) and (\ref{Eq:DC_simple}) cancel with each other. 

\subsection{Comparison between our scheme and the scheme proposed in Ref.~\onlinecite{PhysRevB.90.115435}}
\label{Sec:comp}

Recently, Giovannetti {\it et al.}~\cite{PhysRevB.90.115435} also proposed the downfolding scheme for the electron-phonon coupled systems.  
Here, we compare our scheme with that of Giovannetti {\it et al.}
The main difference is the form of the double counting correction. 
In Ref.~\onlinecite{PhysRevB.90.115435}, the 
phonon-related part of the Hamiltonian,\cite{note_Hamiltonian} which corresponds to Eq.~(\ref{Eq:simple_phonon}) in our case, is given by 
\begin{eqnarray}
\label{Eq:simple_phonon2}
   \hat{{\mathcal H}}     \    &=&  \  
   \sum_{l }    g^{(p)}  x^0_l  n_{l } + 
      \sum_{l } g^{(p)} n_{l }   ( x_{l } -  x^0_l )    \nonumber \\ 
      & +& \   \sum_l  \frac{1}{2} ( \omega^{(p)})^2 ( x_l - x^0_l )^2.  
\end{eqnarray} 
Here, the electron-phonon coupling $g^{(p)}$ is calculated at $x_{l } = x^0_l $, while in our scheme $g^{(p)}$ is calculated at $x_{l } = 0$.  
Furthermore, Giovannetti {\it et al.} includes the term $  \sum_{l }    g^{(p)}  x^0_l  n_{l }$, which represents the deformation of the band 
due to the difference in the equilibrium position between that in the low-energy Hamiltonian and that obtained by the DFT optimization. 
Thus, Giovannetti {\it et al.} also introduce the correction to the electronic part, while our scheme only includes the correction to the phonons. 
Therefore, the form of the Hamiltonian in  Ref.~\onlinecite{PhysRevB.90.115435} is more general than ours. 

If the difference in the equilibrium position is large (i.e., $| x^0_l |$ is large), the band deformation term $  \sum_{l }    g^{(p)}  x^0_l  n_{l }$ would become important. 
In the case of the fullerides, which will be discussed in the next section, we conclude that this effect is small because the equilibrium positions of the undoped and doped ${\rm C}_{60}$ solids 
are very similar, which makes the effect of doping almost rigid band shift. 
Thus, our scheme is well applicable to the fulleride problem. 
However, of course, there exist systems in which this band deformation effect is significant. 
Ref.~\onlinecite{PhysRevB.90.115435} argues that it is important to take into account the band deformation effect in the case of the K-doped picene system,~\cite{Kubozono_nature} because the deformation of the molecule by the doping is not negligible.

To derive the parameters in Eq.~(\ref{Eq:simple_phonon2}), Giovannetti {\it et al.} assume that the electrons couple to 
a single optical phonon, while in our scheme, we can treat all the phonon modes. 
First, Giovannetti {\it et al.} estimate $g^{(p)}$ by calculating the electron-phonon coupling for the undoped picene system. 
Then, they determine $x^0_l$ and $\omega^{(p)}$ such that the mean-field solution of the Hamiltonian recovers the equilibrium positions and the phonon frequencies of the doped system derived within the DFT and DFPT. 

We still lack the methodology to derive the 
Hamiltonian with the form of Eq.~(\ref{Eq:simple_phonon2}) in a totally {\it ab initio} way, 
i.e., without simplifying the electron-phonon coupling or determining $x^0_l$ and $\omega^{(p)}$ in the post processing. 
To realize this, we need to develop an {\it ab initio} structure optimization scheme without the effect of the low-energy electrons. 
We also have to carefully consider the change of the electronic parameters. 
When we derive a model based on the optimized structure without the effect of the low-energy electrons, 
for example, the shape of the Wannier function can be different from that with the fully optimized structure. 
Then, the values of Coulomb interaction parameters can differ from those of the conventional cRPA, which uses the Wannier functions constructed from the fully optimized structure. 
When the band deformation is really severe, we might have to be careful in the choice of the low-energy subspace
since the low-energy band character of the optimized structure without the effect of the low-energy electrons might change from that of the fully optimized structure.  
Therefore, there left many open questions and challenges in 
the derivation of the Hamiltonian 
including the band deformation term [Eq.~(\ref{Eq:simple_phonon2})]. 
Note that, in the situation where this band deformation is important, the cDFPT is also challenged, since 
the current cDFPT does not take account of its effect.


\section{Application}\label{sec:result}

\subsection{Calculation conditions}
We performed the cDFPT calculations~\cite{PhysRevLett.112.027002} for the five different fcc $\A3C60$ systems, namely, 
$\K3C60$, $\Rb3C60$, and $\Cs3C60$ with three different lattice parameters, 
whose properties are summarized in Table~\ref{tab_C60_property}. 
We employed the same lattice constants as those employed in Ref.~\onlinecite{PhysRevB.85.155452}
to evaluate the Coulomb parameters by the cRPA. 
We specify the material by the volume ($V_{\! {{\rm C}_{60}}^{\! \! 3-}}$) occupied per C$_{60}\!^{3-}$ anion in solid.
The most expanded material ($\Cs3C60$ with $V_{\! {{\rm C}_{60}}^{\! \! 3-}}$ = 804 \AA$^3$) is a Mott insulator and the second most expanded system ($\Cs3C60$ with $V_{\! {{\rm C}_{60}}^{\! \! 3-}}$ = 804 \AA$^3$) is on the verge of the metal-insulator transition.~\cite{fcc_CsC60}
The other three materials show a metallic behavior and the superconductivity emerges at low temperature.

\begin{figure}[tb]
\vspace{0cm}
\begin{center}
\includegraphics[width=0.35\textwidth]{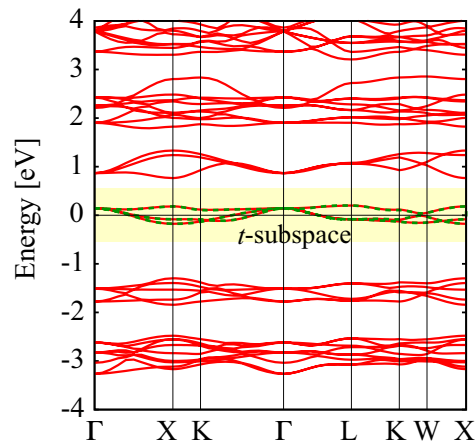}
\caption{(Color online) 
DFT band structure of fcc $\Cs3C60$ with $V_{\! {{\rm C}_{60}}^{\! \! 3-}}$ = 762 \AA$^3$. 
We choose the $t_{1u}$ bands as the $t$-subspace (the yellow shaded region).
The green dotted curves denote the band dispersion derived by Wannier hopping parameters for the $t$-subspace. 
} 
\label{Fig:band_Cs3C60}
\end{center}
\end{figure}

As is already explained in Sec.~\ref{meth:cDFPT}, the implementation of the cDFPT can be done by slightly modifying the existing DFPT program. 
Among the various DFPT codes, in the present study, we modified the one implemented in {\sc quantum espresso} package~\cite{0953-8984-21-39-395502,QEspresso} (see Appendix~\ref{app:implement}).
In the cDFPT calculation, we need to define the low-energy subspace ($t$-subspace). 
Fig.~\ref{Fig:band_Cs3C60} shows the band structure for fcc $\Cs3C60$ with $V_{\! {{\rm C}_{60}}^{\! \! 3-}}$ = 762 \AA$^3$.
Around the Fermi level, there exist the so called $t_{1u}$ bands originating from the three-fold degenerate LUMO orbitals of the $\C60$ molecule. 
The $t_{1u}$ bands are isolated from the other bands. 
As we already mentioned in Sec.~\ref{Sec:intro}, the low-energy physics is governed by the low-energy bands, 
therefore, we choose the $t_{1u}$ bands as target bands.

\begin{table} [tb]
\caption{List of materials employed in the calculation.  
We show the name of compounds, the lattice constant $a$, corresponding volume occupied per C$_{60}\!^{3-}$ anion in solid, 
applied pressure in the experiments, and superconducting transition temperature $\Tc$ or N\'{e}el temperature $T_{\rm N}$.
The listed materials are the same as those of Ref.~\onlinecite{PhysRevB.85.155452}
(in Ref.~\onlinecite{PhysRevB.85.155452}, the Coulomb interaction parameters and the hopping parameters were evaluated).
 }
\vspace{0.0cm}
\begin{center}
\begin{tabular}{@{ \ }c@{\ \ \ \  }  c@{\   \  \ } c @{\  \  \ }c@{\  \ \ } c   @{\  \ \ } c @{\  \  } }
 \hline
 \parbox[c][0.5cm][c]{0cm}{}       &  $a$    &  $V_{\! {{\rm C}_{60}}^{\! \! 3-}}$  &  Pressure   &  $\Tc$ ($T_{\rm N}$)  &  \multirow{2}{*}{ \vspace{-0.15cm} Ref.  }  \\
 \parbox[c][0.5cm][c]{0cm}{}       &  (\AA)  & (\AA$^{3}$)        &  (kbar) &   (K)        &   \\
\hline
 \parbox[c][0.44cm][c]{0cm}{}        fcc $\K3C60$         &  14.240   &    722            &  0      &   19           &    \onlinecite{Zhou19921373}       \\
  \parbox[c][0.44cm][c]{0cm}{}       fcc $\Rb3C60$        &  14.420  &    750            &  0      &   29           &  \onlinecite{Zhou19921373}    \\
  \parbox[c][0.44cm][c]{0cm}{}       fcc $\Cs3C60$     &  14.500  &    762            &  7      &   35           &     \onlinecite{fcc_CsC60}   \\
  \parbox[c][0.44cm][c]{0cm}{}       fcc $\Cs3C60$    &  14.640  &    784            &  2      &   26          &      \onlinecite{fcc_CsC60}    \\
  \parbox[c][0.44cm][c]{0cm}{}       fcc $\Cs3C60$    &  14.762  &    804            &  0      &   (2.2)        &    \onlinecite{fcc_CsC60}      \\ 
\hline
\end{tabular}
\end{center}
\label{tab_C60_property}
\end{table}

The phonon calculations with the cDFPT and the DFPT were performed subsequently to the DFT ground-state calculations.
In the DFT part, we adopted the local density approximation (LDA) with the Perdew-Zunger parameterization.~\cite{PhysRevB.23.5048}
The pseudopotentials for C, K, Rb, and Cs atoms were prepared with the same procedure as in Ref.~\onlinecite{PhysRevB.88.054510} (the Troullier-Martins norm-conserving pseudopotentials~\cite{PhysRevB.43.1993} in the Kleinman-Bylander representation~\cite{PhysRevLett.48.1425}).  
We employed 4$\times$4$\times$4 ${\mathbf k}$ mesh and the cutoff energy of 50 Ry for the wave functions.
With the above conditions, we performed the structure optimization for the materials listed in Table~\ref{tab_C60_property} with fixing the lattice constant and with ignoring the orientational disorder.
In the phonon calculation part, we employed 2$\times$2$\times$2 ${\mathbf q}$ mesh and the Gaussian smearing of 0.025 Ry.

\begin{table*}[tb]  
\caption{Partially renormalized phonon frequencies of $H_g$ modes at $\Gamma$ point calculated by cDFPT method. 
The unit is in cm$^{-1}$ (1 eV = 8065.54 cm$^{-1}$).
In the ideal $I_h$ symmetry (molecular limit), the $H_g$-mode phonon frequencies are five-fold degenerate.  
In the fcc $\A3C60$ systems, they are split into three-fold degenerate and two-fold degenerate frequencies due to the crystal field. 
Thus, we show two frequencies for each $H_g$ mode. 
In the parentheses just after the material names, 
we show the values of $V_{\! {{\rm C}_{60}}^{\! \! 3-}}$ in \AA$^3$. 
}
\vspace{0cm}
\begin{center}
\begin{tabular}{ @{\    } c @{\   \  }  c @{\   } c @{\   } c @{\   }  c @{\  }  c @{\   }  c @{\   }   c  @{\   } }
\hline
 \parbox[c][0.55cm][c]{0cm}{} \multirow{2}{*}{  \vspace{-0.3cm} mode  } &  &   \multicolumn{5}{c}{   frequency [cm$^{-1}$] }       \\
\cline{3-7}
    \parbox[c][0.55cm][c]{0cm}{}       & &     \   \ $\K3C60$ (722) \   \  &    \ \  $\Rb3C60$ (750) \ \    &  \ \     {$\Cs3C60$} (762) \ \   &    $ \ \ \Cs3C60$ (784)  \ \    & \ \    $\Cs3C60$ (804)    \ \  \\ 
\hline
 \parbox[c][0.44cm][c]{0cm}{}    $H_g(1)$  &&  260, 271  &  258, 269  &  259, 278  &  259, 274  &  258, 272    \\
 \parbox[c][0.44cm][c]{0cm}{}    $H_g(2)$  &&  433, 435  &  433, 433  &  434, 436  &  434, 435  &  433, 435    \\
\parbox[c][0.44cm][c]{0cm}{}   $H_g(3)$  &&  706, 708  &  707, 708  &  709, 710  &  709, 710  &  709, 710    \\
\parbox[c][0.44cm][c]{0cm}{}    $H_g(4)$  &&  785, 786  &  785, 787  &  787, 797  &  786, 793  &  785, 791    \\
\parbox[c][0.44cm][c]{0cm}{}    $H_g(5)$  && 1124, 1128 & 1124, 1129 & 1129, 1138 & 1127, 1135 & 1126, 1132   \\
\parbox[c][0.44cm][c]{0cm}{}    $H_g(6)$  && 1282, 1287 & 1282, 1287 & 1292, 1298 & 1288, 1294 & 1286, 1291   \\
\parbox[c][0.44cm][c]{0cm}{}    $H_g(7)$  && 1451, 1455 & 1452, 1455 & 1463, 1466 & 1459, 1461 & 1457, 1459   \\
\parbox[c][0.44cm][c]{0cm}{}    $H_g(8)$  && 1563, 1564 & 1563, 1565 & 1573, 1573 & 1569, 1570 & 1567, 1568   \\   
\hline
\end{tabular} 
\end{center}
\label{tab_freq_partial}
\end{table*}

\begin{table*}[tb]  
\caption{Fully renormalized phonon frequencies of $H_g$ modes at $\Gamma$ point calculated by conventional
DFPT method. 
The unit is in cm$^{-1}$ (1 eV = 8065.54 cm$^{-1}$).
The splitting of the frequencies of each $H_g$ mode is due to the crystal field.
We show the values of $V_{\! {{\rm C}_{60}}^{\! \! 3-}}$ in \AA$^3$ in the parentheses just after the material names. 
For comparison, we also show the experimentally observed phonon frequencies in $\K3C60$.~\cite{PhysRevB.45.10838} 
}
\vspace{0cm}
{\small 
\begin{center}
\begin{tabular}{  c@{\  \  \ \  }  c@{\     } c@{ \     }  c @{\  }  c @{\   }  c  @{\  }  c  }
\hline
 \parbox[c][0.55cm][c]{0cm}{} \multirow{2}{*}{  \vspace{-0.3cm} mode    }  &   \multicolumn{6}{c}{frequency [cm$^{-1}$]}  \\
\cline{2-7}
 \parbox[c][0.55cm][c]{0cm}{}   & \ \  $\K3C60$ (722) \ \ &  \ \ $\Rb3C60$ (750)  \ \  &   \ \ {$\Cs3C60$} (762)  \ \ & \ \  $\Cs3C60$ (784) \ \  &  \ \ $\Cs3C60$ (804) \ \  & \ \ $^a$$\K3C60$ (expt.) \ \ \\ 
\hline
 \parbox[c][0.44cm][c]{0cm}{}  $H_g(1)$  &  257, 268  &  255, 267  &  256, 277  &  255, 273  &  255, 271   & 271\\
 \parbox[c][0.44cm][c]{0cm}{}  $H_g(2)$  &  423, 425  &  422, 423  &  422, 425  &  421, 424  &  420, 423   & 431\\
 \parbox[c][0.44cm][c]{0cm}{}  $H_g(3)$  &  683, 686  &  684, 686  &  686, 688  &  686, 688  &  686, 687   & 723\\
 \parbox[c][0.44cm][c]{0cm}{}  $H_g(4)$  &  777, 778  &  777, 778  &  780, 788  &  779, 785  &  778, 782   & $\ldots$ \\
 \parbox[c][0.44cm][c]{0cm}{}  $H_g(5)$  & 1110, 1114 & 1110, 1114 & 1116, 1125 & 1113, 1121 & 1112, 1118 & $\ldots$  \\
 \parbox[c][0.44cm][c]{0cm}{}  $H_g(6)$  & 1267, 1273 & 1267, 1272 & 1277, 1283 & 1273, 1278 & 1270, 1275 &  $\ldots$ \\
 \parbox[c][0.44cm][c]{0cm}{}  $H_g(7)$  & 1402, 1407 & 1403, 1405 & 1415, 1415 & 1410, 1410 & 1406, 1407 & 1408 \\
 \parbox[c][0.44cm][c]{0cm}{}  $H_g(8)$  & 1531, 1536 & 1531, 1535 & 1541, 1544 & 1537, 1540 & 1535, 1538 &  1547\\               
\hline
\end{tabular}
\end{center}
\vspace{-0.3cm}
 $^a$ Raman scattering measurement, Ref.~\onlinecite{PhysRevB.45.10838}
}
\label{tab_freq_full}
\end{table*}

\subsection{Phonon frequencies}

In the alkali-doped fullerides, it has been shown that the dominant electron-phonon coupling is coming from the intramolecular vibration.~\cite{RevModPhys.69.575,PhysRevB.46.12088,PhysRevB.48.7651,Pickett_inter,Ebbesen1992163,Burk19942493,PhysRevLett.72.3706}
When we consider the isolated $\C60$ molecule, 
only the intramolecular phonon modes with the $A_g$ and $H_g$ symmetries have finite electron-phonon couplings to the $t_{1u}$ electrons.~\cite{VARMA15111991,PhysRevB.44.12106}
This is because the $\C60$ molecule has extremely high symmetry ($I_h$ symmetry) and the coupling to the other modes are forbidden due to the symmetry reason.~\cite{RevModPhys.69.575}
This property also holds well in the $\C60$ solids.  
In particular, the coupling to the Jahn-Teller phonon (so called $H_g$ modes) is argued to be crucial to the superconductivity.~\cite{nomura_science_advances,Gun_book}

\begin{figure}[tbp]
\vspace{0cm}
\begin{center}
\includegraphics[width=0.44\textwidth]{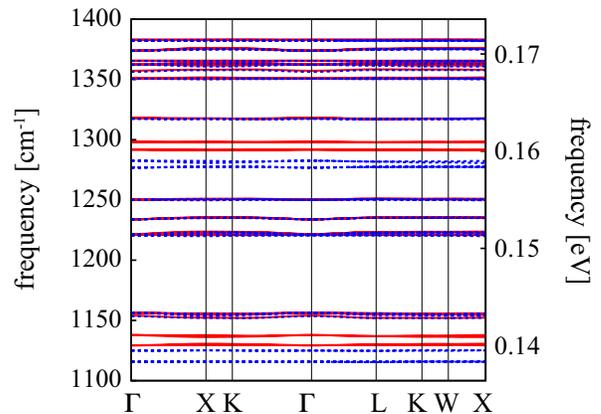}
\caption{(Color online) 
Phonon dispersion of fcc $\Cs3C60$ with $V_{\! {{\rm C}_{60}}^{\! \! 3-}} = 762$ \AA$^3$.
For the sake of visibility, we restrict the frequency range to 1100-1400 cm$^{-1}$. 
The red solid (blue dotted) curves indicate partially (fully) renormalized frequencies 
calculated by the cDFPT (conventional DFPT). 
} 
\label{Fig:ph_disp}
\end{center}
\end{figure}

Table~\ref{tab_freq_partial} summarizes our calculated partially renormalized phonon frequencies ($\omega^{(p)}$'s) of the $H_g$ modes at $\Gamma$ point. 
Due to the crystal field, the frequencies of the $H_g$-mode are split into two.
The high phonon frequencies up to $\sim \! 1600 {\rm cm}^{-1}$ \! ($\sim \! 0.2$ eV) can be ascribed to the stiff C-C bonds and the lightness of the carbon atoms.  
Furthermore, the intramolecular nature of the modes leads to the following features:  
The $H_g$ phonon modes have little dispersion (see Fig.~\ref{Fig:ph_disp}). 
The material dependence of the frequencies is weak. 

Note that these partially renormalized frequencies $\omega^{(p)}$'s are the inputs for the low-energy solvers and thus can not be directly compared with the experimentally observed frequencies. 
To compare with the experiments, we have to include the effect of the $t$-subspace electrons and calculate the fully renormalized phonon frequencies ($\omega^{(f)}$'s).
In general, a stronger coupling between the $t$-subspace electrons and the phonons leads to a larger difference between 
$\omega^{(p)}$'s and $\omega^{(f)}$'s.~\cite{PhysRevB.84.184531} 
In the case of the alkali-doped fullerides, 
the electron-phonon coupling of the individual mode is not large, while the accumulation of the contributions leads to 
the total electron-phonon coupling of $\lambda \sim 0.5$-1.0.~\cite{Gun_book,PhysRevB.82.245409,PhysRevB.81.073106,PhysRevB.84.155104} 
Therefore, we do not expect a large difference between  $\omega^{(p)}$'s and $\omega^{(f)}$'s. 

In Table~\ref{tab_freq_full}, we list the fully renormalized phonon frequencies of the $H_g$ modes at $\Gamma$ point computed by the DFPT. 
By comparing them with the partially renormalized values in Table~\ref{tab_freq_partial}, 
we see the softening of the frequencies. 
This is because the phonons are dressed by the coupling between the phonons and the $t_{1u}$ electrons. 
In other words, the phonons acquire the self-energy associated with the $t$-subspace electrons.
However, as is expected (see the discussion above), 
the difference is small: 
The absolute difference is at most $\sim 50$ cm$^{-1}$. 
If we consider the ratio $\omega^{(f)}/ \omega^{(p)}$, it exceeds 0.95, i.e., the difference is less than 5 \%. 
Even when we accurately treat the $t$-subspace processes beyond the DFPT level by the model calculation method, 
the $t$-subspace renormalization effects would remain small. 
Then, we can expect that the conventional DFPT calculations give reasonable estimates of the phonon frequencies. 
Indeed, the fully-renormalized frequencies in Table~\ref{tab_freq_full} agree well with the experimental data.~\cite{Bethune1991181,PhysRevB.45.10838}

Figure~\ref{Fig:ph_disp} shows both the partially (red) and fully (blue) renormalized phonon frequencies 
between 1100 cm$^{-1}$ and 1400 cm$^{-1}$
for fcc $\Cs3C60$ with $V_{\! {{\rm C}_{60}}^{\! \! 3-}} = 762$ \AA$^3$.
Several intramolecular modes including $H_g$ modes [$H_g(5)$ and $H_g(6)$] and the non-$H_g$ modes 
exist in this frequency range.
While they are common in that they have little dispersions, 
we see a clear difference between the $H_g$ modes and the others in the way of the softening:   
The non-$H_g$ modes do not couple to the $t_{1u}$ electrons.~\cite{note_Ag} 
Hence, their frequencies are not affected by the inclusion of the $t$-subspace renormalization effects. 
As a result, the blue dotted curves ($\omega^{(f)}$) are on top of the red solid curves ($\omega^{(p)}$) for the non-$H_g$ modes. 
On the other hand, the frequencies for the $H_g$ modes are renormalized by a few percent. 
Indeed, the red and blue curves are located at different positions for the $H_g$ modes 
(see the frequency regions 1100-1150 and 1260-1300 cm$^{-1}$).

\subsection{Phonon-mediated effective interactions between the low-energy electrons}

\begin{table*}[tb]  
\caption{Material dependence of static part ($\omega_n \! = \! 0$) of the effective intramolecular interactions mediated by phonons.
The values in the parentheses just after the material names denote $V_{\! {{\rm C}_{60}}^{\! \! 3-}}$ in \AA$^3$. }
\vspace{0cm}
\begin{center}
\begin{tabular}{ @{\    } c @{\   \  }  c @{\   } c @{\   } c @{\   }  c @{\  }  c @{\   }  c @{\   }   c  @{\   } }
\hline
 \parbox[c][0.55cm][c]{0cm}{} \multirow{2}{*}{ \vspace{-0.3cm} type of int.   } &  &   \multicolumn{5}{c}{   interaction [meV] }       \\
\cline{3-7}
    \parbox[c][0.55cm][c]{0cm}{}       & &     \   \  \ $\K3C60$ (722) \   \  &    \ \  $\Rb3C60$ (750) \ \    &  \ \     {$\Cs3C60$} (762) \ \   &    $ \ \ \Cs3C60$ (784)  \ \    & \ \    $\Cs3C60$ (804)    \ \  \\ 
\hline  
 \parbox[c][0.51cm][c]{0cm}{} \  $U^{(p)}_{\rm ph}(0)$  &&  $-152$ &  $-142$ &  $-114$ &  $-124$   &  $-134$    \\
 \parbox[c][0.51cm][c]{0cm}{} \  $U'^{(p)}_{\rm ph}(0)$  &&  $-53$  &  $-42$   &   $-13$    &  $-22$     &  $-31$  \\
 \parbox[c][0.51cm][c]{0cm}{} \  $J^{(p)}_{\rm ph}(0)$  &&  $-50$    &  $-51$   &  $-51$    &  $-51$     &  $-52$    \\
\hline  
 \parbox[c][0.51cm][c]{0cm}{} \  $U^{(f)}_{\rm ph}(0)$  &&  $-73$    &  $-74$   &  $-73$    &  $-74$    &  $-75$    \\
 \parbox[c][0.51cm][c]{0cm}{} \  $U'^{(f)}_{\rm ph}(0)$  &&  \phantom{$-$}$28$   &   \phantom{$-$}$29$    &   \phantom{$-$}$30$   &  \phantom{$-$}$31$      &  \phantom{$-$}$31$   \\
 \parbox[c][0.51cm][c]{0cm}{} \  $J^{(f)}_{\rm ph}(0)$  &&  $-51$   &  $-52$    &  $-52$   &  $-52$      &  $-53$  \\
\hline
\end{tabular} 
\end{center}
\label{tab_ph_int}
\end{table*}

\begin{figure}[tb]
\vspace{0cm}
\begin{center}
\includegraphics[width=0.24\textwidth]{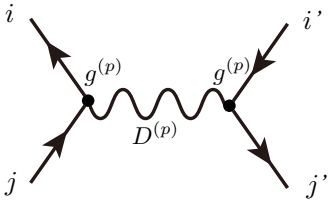}
\caption{
Feynman diagram for phonon-mediated interaction between electrons. 
Solid lines with arrows, wavy lines, and dots represent the electron propagator, the phonon propagator, and the electron-phonon coupling, respectively. 
} 
\label{Fig:PH_diagram}
\end{center}
\end{figure}

If we write down the partition function for the electron-phonon coupled Hamiltonian in Eq.~(\ref{H_ele_ph_system})
in the coherent state path-integral formalism, we find that we have at most quadratic term for the phonon fields. 
Then, we can integrate out the phonon degrees of freedom analytically. 
This results in an electronic model with the additional electron-electron interaction mediated by the phonons  (Fig.~\ref{Fig:PH_diagram}),~\cite{PhysRevB.76.035116} whose onsite (= intramolecular) part $V^{(p)}_{ij,i'j'}( i\omega_n)$ is given by~\cite{PhysRevLett.112.027002}  
  \begin{eqnarray}
  \label{Eq.ph_el_el_interaction}
V^{(p)}_{ij,i'j'}   (i\omega_n)  &=& 
  \frac{1}{N_{\q1}}  \sum_{\q1\nu}   \ \! 
     \tilde{g}_{ij}^{(p)}  ({\bf q}, \nu)   \ \! 
     D^{(p)}_{\bf q, \nu} (i\omega_n)  \ \! 
     \tilde{g}_{j'i'}^{(p)\ast}  ({\bf q}, \nu)  \nonumber \\   
 &=& -  \frac{1}{N_{\q1}}  \sum_{\q1\nu}  \ \! 
     \tilde{g}_{ij}^{(p)}  ({\bf q}, \nu)  \ \!  \frac{2  \omega^{(p)}_{{\bf q} \nu} }  { \omega_n^2  +  ( \omega^{(p)}_{{\bf q} \nu})^2 }  \ \! 
     \tilde{g}_{j'i'}^{(p)\ast}  ({\bf q}, \nu),  \nonumber \\
     \label{Eq.Vpartial_result_sec} 
\end{eqnarray}  
where $N_{\q1}$ is the number of $\bf q$-mesh and  $\omega_n$ is  the bosonic Matsubara frequency $\omega_n = 2 \pi n T$
 with the temperature $T$.\cite{note_ph_sum}
Here, $\tilde{g}^{(p)}$'s are given by
\begin{eqnarray}
\tilde{g}_{ij}^{(p)}  ({\bf q}, \nu)  =  \frac{1}{N_{\k1}} \sum_{\k1}  g^{(p) \nu}_{ij}(\k1,\q1).  
\label{Eq.tilde_g_interface}
\end{eqnarray}
Here, the partially-screened electron-phonon coupling $g^{(p)}$ is used to calculate the phonon-mediated interactions. 
In Appendix~\ref{app:vertex}, we discuss that the vertex correction for $g^{(p)}$ is small, which makes the estimate of the phonon-mediated interactions 
without the vertex correction reliable.

The phonon-mediated interactions $V^{(p)}_{ij,i'j'}   (i\omega_n)$ are dynamical interactions, which vanish in high frequency limit ($\omega_n \rightarrow \infty $). 
We call the intraorbital density-density-type interaction, interorbital density-density-type interaction, 
exchange-type interaction $U^{(p)}_{\rm ph} (i\omega_n)$, $U'^{(p)}_{\rm ph}(i\omega_n)$, and $J^{(p)}_{\rm ph}(i\omega_n)$, respectively, i.e.,  
\begin{eqnarray}
\hspace{0.4cm}   U^{(p)}_{\rm ph} (i\omega_n) \ &=&  \ V^{(p)}_{ii,ii}   (i\omega_n),  \nonumber \\
  U'^{(p)}_{\rm ph} (i\omega_n) \ &=& \ V^{(p)}_{ii,jj}   (i\omega_n), \nonumber \\ 
   J^{(p)}_{\rm ph} (i\omega_n) \ &=& \ V^{(p)}_{ij,ji}   (i\omega_n) =  V^{(p)}_{ij,ij}   (i\omega_n) 
\end{eqnarray}
with $i\neq j$. 
We also define the fully screened quantities $U^{(f)}_{\rm ph} (i\omega_n)$, $U'^{(f)}_{\rm ph}(i\omega_n)$, and $J^{(f)}_{\rm ph}(i\omega_n)$ in the same way, i.e., 
$ U^{(f)}_{\rm ph} (i\omega_n) = V^{(f)}_{ii,ii}   (i\omega_n)$, 
  $U'^{(f)}_{\rm ph} (i\omega_n) = V^{(f)}_{ii,jj}   (i\omega_n)$,  and 
   $J^{(f)}_{\rm ph} (i\omega_n) = V^{(f)}_{ij,ji}   (i\omega_n) =  V^{(f)}_{ij,ij}   (i\omega_n) $.
We find that, because of the high symmetry of the $t_{1u}$ orbitals, the values of  $U^{(p,f)}_{\rm ph} (i\omega_n)$, $U'^{(p,f)}_{\rm ph}(i\omega_n)$, and $J^{(p,f)}_{\rm ph}(i\omega_n)$
do not depend on orbital.

\begin{figure*}[tbp]
\vspace{0cm}
\begin{center}
\includegraphics[width=0.78\textwidth]{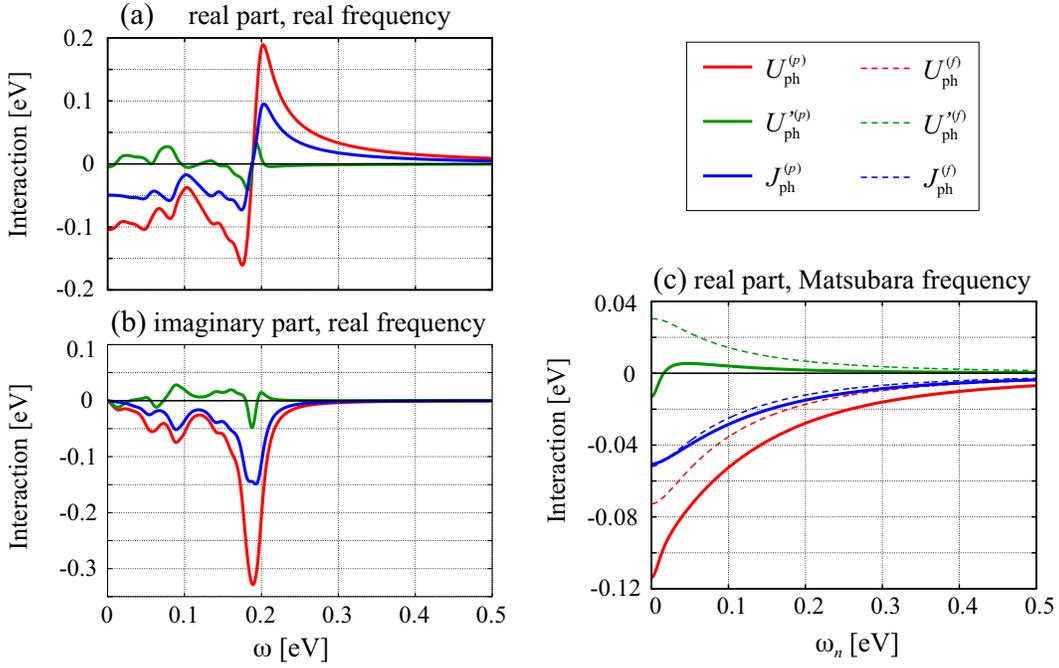}
\caption{(Color online) 
Frequency dependence of phonon-mediated interactions for fcc $\Cs3C60$ with $V_{\! {{\rm C}_{60}}^{\! \! 3-}} = 762$ \AA$^3$ along the real frequency [panels (a) and (b)], and along the Matsubara axis [panel (c)].  
(a) [(b)] Real [imaginary] part of $U^{(p)}_{\rm ph}(\omega)$, $U'^{(p)}_{\rm ph}(\omega)$ and $J^{(p)}_{\rm ph}(\omega)$. 
The frequency dependence of these quantities are calculated at $\omega + i \eta$ with $\eta = 0.01$ eV.  
(c) Real part of $U^{(p)}_{\rm ph}(i \omega_n)$, $U'^{(p)}_{\rm ph}(i \omega_n)$, $J^{(p)}_{\rm ph}(i \omega_n)$, 
$U^{(f)}_{\rm ph}(i \omega_n)$, $U'^{(f)}_{\rm ph}( i \omega_n)$ and $J^{(f)}_{\rm ph}( i \omega_n)$. 
The imaginary part is always zero along the Matsubara axis. 
} 
\label{Fig:freq_dep}
\end{center}
\end{figure*}

Table~\ref{tab_ph_int} summarizes the values of the static parts of these interaction ($\omega_n \!=\! 0$). 
We find that the relation $U'^{(p,f)}_{\rm ph} (0) \! \sim \! U^{(p,f)}_{\rm ph} (0) \! -\!   2 J^{(p,f)}_{\rm ph}(0)$ well holds, 
which also holds for finite frequency (see Fig.~\ref{Fig:freq_dep}). 
We first discuss the partially renormalized interactions. 
The negative values of $U^{(p)}_{\rm ph} (0)$, $U'^{(p)}_{\rm ph} (0)$, and $J^{(p)}_{\rm ph}(0)$  
indicate that the interactions are attractive at $\omega_n=0$. 
Therefore, they will compete with the repulsive onsite Coulomb interactions. 
As for the density-density channel,
since the intramolecular Coulomb repulsion (the Hubbard $U$) for the $t_{1u}$ electrons is estimated to be on the order of $\sim \! 1$ eV,~\cite{PhysRevB.85.155452} 
the repulsive Coulomb interaction dominates over the phonon-mediated attraction. 
However, remarkably, the situation changes for the exchange-type interaction:
the absolute values of $| J^{(p)}_{\rm ph}(0) | \! \sim \! 0.05$ eV is larger than those of the Hund's coupling 
$J \! \sim \! 0.035$ eV.~\cite{PhysRevB.85.155452} Therefore, in the fullerides, an effectively negative exchange interaction is realized.~\cite{nomura_science_advances}
This is in constant with e.g., the case of LaFeAsO  (the first discovered iron-based superconductor~\cite{doi:10.1021/ja800073m}), where the Hund's coupling is as large as $\sim 0.5$ eV~\cite{doi:10.1143/JPSJ.77.093711,doi:10.1143/JPSJ.79.044705} and the phonon-mediated exchange interaction $J^{(p)}_{\rm ph}(0) \sim -0.02$ eV gives only a minor correction.~\cite{PhysRevLett.112.027002} 
The unusual competition of the Hund's coupling and the phonon-mediated interactions can be ascribed, mainly, to the following two reasons.~\cite{nomura_science_advances} 
One is the molecular nature of the maximally localized Wannier orbitals.  Then, the sizes of the Wannier orbitals become larger than those of atomic-orbital-like Wannier functions, which results in a smaller Hund's coupling. 
The other is the enhancement of the negative $J^{(p)}_{\rm ph}(0)$ due to the strong couplings between the Jahn-Teller modes and the $t_{1u}$ electrons. 
The Jahn-Teller $H_g$ modes give the non-density-type electron-phonon coupling, which contribute to $J^{(p)}_{\rm ph}(0)$.~\cite{VARMA15111991,PhysRevB.51.3493,PhysRevB.44.12106}   
Note that the non-Jahn-Teller $A_g$ modes do not contribute, since the couplings of the $A_g$ modes are of density-type.  

As for the material dependence, 
while that of $J^{(p)}_{\rm ph}(0)$ is small, 
we see discernible material dependence in $U^{(p)}_{\rm ph} (0)$ and $U'^{(p)}_{\rm ph} (0)$. 
We identify the origin of the material dependence to be the vibration modes of the alkali ions at the tetrahedral sites. 
It is reasonable that they give a material-dependent contribution 
as the distances between the ${{\rm C}_{60}}^{\! \! 3-}$ anions and/or the alkali cations change. 
Indeed, if we compute $U^{(p)}_{\rm ph}(0)$ and $U'^{(p)}_{\rm ph}(0)$ for the five materials with excluding the alkali-ion contributions
(in this case, the values become the sum of the contribution from the intramolecular phonons), 
the results for $U^{(p)}_{\rm ph}(0)$ [$U'^{(p)}_{\rm ph}(0)$] are, in ascending order of $V_{\! {{\rm C}_{60}}^{\! \! 3-}}$, 
$-89$ [10], $-91$ [9], $-91$ [9], $-93$ [8], and $-95$ [8] meV.  
As is clear, they have much less material dependence than those with the alkali-ion contributions, 
which is natural because we can expect that the intramolecular phonons have little material dependence. 
We find that the alkali-ion modes couple to the total density of the $t_{1u}$ electrons, i.e., they couple to the density of the individual orbital with almost the same amplitudes ($\tilde{g}^{(p)}_{11} \! \simeq \!  \tilde{g}^{(p)}_{22} \! \simeq \! \tilde{g}^{(p)}_{33}$).  
Thus, it does not contribute to $J^{(p)}_{\rm ph}$. 
The contribution to $J^{(p)}_{\rm ph}(0)$ originates from the intramolecular Jahn-Teller coupling (coupling to the $H_g$ modes). 
 Therefore, we see little material dependence in $J^{(p)}_{\rm ph} (0)$. 
Since the electron-phonon coupling of the alkali-ion modes are of density-type, the alkali-ion mode contribution is efficiently screened by the $t_{1u}$ electrons, which leads to a minor role of the alkali-ion modes in the superconductivity. 
As a result, as we will see below, the dominant contribution to the fully renormalized interactions comes from the 
intramolecular phonons, which is consistent with the previous studies.~\cite{PhysRevLett.69.957,PhysRevB.48.7651}

We can compute the fully-screened phonon-mediated onsite interactions using Eq.~(\ref{Eq.Vpartial_result_sec})
by replacing the partially renormalized quantities with the fully renormalized quantities. 
We list the values of their static part ($\omega_n \! = \! 0$)  in Table~\ref{tab_ph_int}.
We find that the magnitudes of density-density type interactions, $U^{\prime (f)}_{\rm ph} (0)$ and $U^{(f)}_{\rm ph} (0)$, 
differ substantially from those of the partially renormalized ones, $U^{\prime (p)}_{\rm ph} (0)$ and $U^{(p)}_{\rm ph} (0)$. 
On the other hand, the values of $J^{(f)}_{\rm ph}(0)$ are almost unchanged from those of $J^{(p)}_{\rm ph} (0)$.
This different behavior between $U_{\rm ph}, U'_{\rm ph}$ and $J_{\rm ph}$ can be understood as follows. 
The $t_{1u}$ electrons efficiently screen the non-Jahn-Teller type electron-phonon coupling, while the Jahn-Teller type coupling not. 
The former contributes to  $U_{\rm ph}$ and $U'_{\rm ph}$. Therefore, the difference between the partially and fully renormalized quantities is substantial.
On the other hand, only the Jahn-Teller phonon contributes to $J_{\rm ph}$. Therefore, we have little difference between the partially and fully renormalized quantities. 
As we discuss above, the alkali-ion mode contribution becomes small in the fully renormalized quantities and the intramolecular $H_g$ mode contribution becomes dominant (intramolecular $A_g$ mode contribution is also screened because $A_g$ mode couples to the total density of the $t_{1u}$ electrons), which makes the material dependence of $U^{(f)}_{\rm ph} (0)$, $U'^{(f)}_{\rm ph}(0)$, and $J^{(f)}_{\rm ph}(0)$ small. 

When we consider the contribution from $H_g$ modes in the molecular limit,~\cite{VARMA15111991,PhysRevB.51.3493,PhysRevB.44.12106}
we can show that the relation $U^{\prime (f)}_{\rm ph} (i \omega_n) = - U^{(f)}_{\rm ph} (i \omega_n ) / 2$ holds. 
Since, in reality, we have a small contribution from the other modes such as $A_g$ and the alkali-ion modes, 
the above relation does not exactly hold. 
However, this naturally explains why the interorbital interactions become repulsive ($U^{\prime (f)}_{\rm ph} (0) > 0 $).

We finally discuss the frequency dependence of the phonon-mediated interactions. 
The frequency dependences for fcc $\Cs3C60$ with $V_{\! {{\rm C}_{60}}^{\! \! 3-}} = 762$ \AA$^3$
on the real frequency axis are shown in Figs.~\ref{Fig:freq_dep}(a) and (b), where 
the panel (a) [(b)] shows the real [imaginary] part of the phonon-mediated interactions. 
Since the frequencies of the intramolecular phonons lies up to $\sim 0.2$ eV, there exist significant structures below $\sim 0.2$ eV. 
${\rm Im} \ U^{(p)}_{\rm ph} (\omega)$ and ${\rm  Im}\  J^{(p)}_{\rm ph} (\omega)$ are always negative. 
On the other hand, the ${\rm Im} \ U'^{(p)}_{\rm ph} (\omega)$ can be both negative and positive. 
This is because the contribution from the non-Jahn-Teller and the Jahn-Teller ($H_g$) phonons coexist
[the former (latter) gives negative (positive) contribution].
Note that both the non-Jahn-Teller and Jahn-Teller phonons give negative contribution to ${\rm Im} \ U^{(p)}_{\rm ph} (\omega)$ and ${\rm Im} \ J^{(p)}_{\rm ph} (\omega)$. 

We also show the frequency dependence along the Matsubara frequencies in Fig.~\ref{Fig:freq_dep}(c), where we also plot the frequency dependence of the fully screened interactions. 
Again, the non-monotonic behavior in $U'^{(p)}_{\rm ph} (i \omega_n)$ can be ascribed to the coexistence of the attractive (non-Jahn-Teller) and repulsive (Jahn-Teller) contributions.
Because the contribution from the non-Jahn-Teller phonons becomes small in the fully screened interactions, the frequency dependence of 
 $U'^{(f)}_{\rm ph} (i \omega_n)$ becomes monotonic. 
 Since only the Jahn-Teller modes, which is poorly screened by the $t_{1u}$ electrons, contribute to $J_{\rm ph}$, 
 we have small difference between $J^{(p)}_{\rm ph} (i \omega_n)$ and $J^{(f)}_{\rm ph} (i \omega_n)$.
Finally, we note that the relation $U'^{(p,f)}_{\rm ph}  = U^{(p,f)}_{\rm ph} -  2 J^{(p,f)}_{\rm ph} $ well holds along both the real and imaginary frequency axes.

\section{Conclusion and outlook}\label{sec:summary}

In this paper, we have presented a detailed explanation on the newly developed {\it ab initio} downfolding scheme for the electron-phonon coupled system, cDFPT. 
With the cDFPT, we can calculate the partially renormalized phonon frequencies and electron-phonon coupling, which is used as the parameters in the effective low-energy Hamiltonian. 
We have shown that the cDFPT scheme can be easily implemented by the slight modification of the conventional DFPT scheme. 
 
We have applied the cDFPT scheme to the alkali-doped fullerides. 
By excluding the $t$-subspace renormalization effect, 
we have seen the hardening of the frequencies of the phonon modes which couple to the $t$-subspace electrons. 
We have also discussed the difference between the partially and fully screened phonon-mediated interactions. 
In the partially screened interactions, the non-Jahn-Teller phonons give substantial contributions. 
However, in the fully screened screened interactions, the contribution from the non-Jahn-Teller modes becomes small because it is efficiently screened by the $t_{1u}$ electrons. 
Then, the Jahn-Teller phonons give the dominant contributions to the fully screened interactions. 

In this paper, we have focused on the alkali-doped fullerides. 
However, in principle, the cDFPT is applicable to other materials in which phonons play a crucial role.
These applications remain as interesting and important future issues. 
There also remain challenges in the development side: 
As discussed in Sec.~\ref{Sec:comp}, the current cDFPT is not applicable when 
the equilibrium positions of the ions change drastically by the coupling to the low-energy electrons. 
It is also challenged when there exists strong anharmonicity in the system.  
These are important open questions in the downfolding for electron-phonon coupled systems.

\begin{acknowledgements}
\vspace{-0.4cm}
We would like to thank Kazuma Nakamura, Shiro Sakai, Massimo Capone, Ryosuke Akashi, Takahiro Ohgoe, Terumasa Tadano, Masatoshi Imada, Atsushi Fujimori, Atsushi Oshiyama, and Yoshihiro Iwasa for fruitful discussions. 
Y.N. is supported by Grant-in-Aid for JSPS Fellows (No. 12J08652) from Japan Society for the Promotion of Science (JSPS), Japan.
\end{acknowledgements}

\vspace{0.5cm}

\appendix
\section{Equivalence of Eqs.~(\ref{Eq:delpsi_metal}) and (\ref{eq:DFPT_metal})}\label{app.omit_unocc}
Here, we show that Eqs.~(\ref{Eq:delpsi_metal})  and (\ref{eq:DFPT_metal}) indeed give the same solution. 
When we write Eq.~(\ref{eq:DFPT_metal}) as  
\begin{eqnarray} \label{appeq:DFPT_metal}
\underbrace{\bigl( \Hscf1 + Q - \varepsilon_n \bigr) }_{\text{{\large $A$}}} 
 \underbrace{\bigl | \Delta \psi_n \bigr \rangle}_{\text{{\large ${\mathbf x}$}}} 
 = \underbrace{ - \bigl (\thetan1 - P_n \bigr ) \dvscf1 \bigl | \psi_n \bigr \rangle }_{\text{{\large ${\mathbf y}$}}},  \nonumber \\
\end{eqnarray}
the $A$ matrix is given, in the Bloch basis (note that, in the case of {\sc quantum espresso}, the plane basis is used in the actual calculation), by 
\begin{eqnarray}
\label{appeq:A}
A = 
\left(
\begin{array}{cccc}
     \   \varepsilon_{1} \!+\! \alpha_1 \!-\! \varepsilon_{n}    \  &  & & \text{{\LARGE{0}}}\\
  &   \  \varepsilon_{2} \!+\! \alpha_2 \!-\! \varepsilon_{n}   \  & &  \\
    &       & \ \  \ddots \ \  &  \\
 \text{{\LARGE{0}}}     & &&   \   \varepsilon_{M} \!+\! \alpha_M \!-\! \varepsilon_{n}  \ 
\end{array} \right),   \nonumber \\
\end{eqnarray}
where $M$ is the size of the basis set to describe the Bloch states.
$\thetan1 - P_n$ on the r.h.s of Eq.~(\ref{eq:DFPT_metal}) is rewritten as 
\begin{widetext}
\begin{eqnarray}
\label{appeq:theta}
  \thetan1 - P_n \ &=& \ 
 \sum_m \left[   \thetan1   \bigl( 1- \thetanm1 \bigr) - \thetam1 \thetamn1 - 
 \alpha_m \frac{ \thetan1 - \thetam1 } {\varepsilon_n - \varepsilon_m} \thetamn1  \right] 
   \bigl | \psi_m \bigr \rangle  \bigl \langle \psi_m \bigr |  \nonumber \\ 
  &=&  \  \sum_m \left[    \bigl( \thetan1   - \thetam1 \bigr ) \thetamn1 - 
 \alpha_m \frac{ \thetan1 - \thetam1 } {\varepsilon_n - \varepsilon_m} \thetamn1  \right] 
   \bigl | \psi_m \bigr \rangle  \bigl \langle \psi_m \bigr |  \nonumber \\ 
     &=& \ 
-   \sum_m \biggl[   \frac{ \thetan1 - \thetam1 } {\varepsilon_n - \varepsilon_m} \thetamn1   \bigl (  \varepsilon_m  + \alpha_m - \varepsilon_n  \bigr) \biggr] 
   \bigl | \psi_m \bigr \rangle  \bigl \langle \psi_m \bigr | .   
\end{eqnarray}
\end{widetext}
With Eqs.~(\ref{appeq:DFPT_metal}), (\ref{appeq:A}), and (\ref{appeq:theta}), we can show that $\bigl | \Delta \psi_n \bigr \rangle$ is given by  
\begin{eqnarray} 
\bigl | \Delta \psi_n \bigr \rangle \  &=&  \  A^{-1} {\mathbf y}   \nonumber \\ 
  \  &=& \ 
   \sum_m \frac{ \thetan1 - \thetam1 } {\varepsilon_n - \varepsilon_m} \thetamn1 
    \bigl | \psi_m \bigr \rangle  \bigl \langle \psi_m \bigr | 
     \dvscf1 \bigl | \psi_n \bigr \rangle, \nonumber \\ 
 \end{eqnarray}
 which is nothing but a proof that Eq.~(\ref{eq:DFPT_metal}) gives the same result as that of Eq.~(\ref{Eq:delpsi_metal}).

\section{Comparison between ${\rm \bf c}$DFPT and ${\rm \bf c}$RPA}
\label{App:comp}
Here, we 
compare
the present cDFPT with
the cRPA.~\cite{PhysRevB.70.195104} 
In the cRPA, which derives the effective electron-electron interactions in the low-energy model, 
we calculate the partially screened Coulomb interaction as~\cite{PhysRevB.70.195104} 
\begin{eqnarray}
\label{ap_Eq.Wp}
W^{(p)}=\left(1-v\chi^0_r\right)^{-1} v.
\end{eqnarray}
The fully screened Coulomb interaction is obtained by further taking into account the $t$-subspace screening effect: 
\begin{eqnarray}
\label{ap_Eq.Wf}
W^{(f)}=\left(1-W^{(p)}\chi^0_t\right)^{-1}W^{(p)}.  
\end{eqnarray}
One can see that Eqs.~(\ref{ap_Eq.Wp}) and (\ref{ap_Eq.Wf}) have the same structure as 
that of the screened electron-phonon coupling [Eqs.~(\ref{Eq:gp}) and (\ref{Eq:gf})]. 
Both the cRPA and cDFPT methods relies on the same kind of decomposition of the screening processes. 
In both cases, we calculate the partially screened quantities, which is to be used in the low-energy Hamiltonian.

\section{Practical implementation in the case of {\sc quantum espresso}}
\label{app:implement}

Here, we provide an example how we modify a source code. 
In the DFPT implemented in the version 4.3.1 of {\sc quantum espresso},~\cite{0953-8984-21-39-395502,QEspresso}
the $\beta_{n,m}$ parameters are defined in ``orthogonalize.f90", which exists in ``PH" folder. 
In Ref.~\onlinecite{DFPT_change}, we distribute a modified ``orthogonalize.f90" under the GNU General Public License.~\cite{GPL}
     
\section{Confirmation of the equality $\Sigma = \Sigma_t + \Sigma_r$ in Sec.~\ref{sec_relate_p_f}}
\label{app:proof_S}
Here, we show that the equality $\Sigma = \Sigma_t + \Sigma_r$ in Sec.~\ref{sec_relate_p_f} indeed holds. 
In principle, the self-energy $\Sigma$, the electron-phonon coupling $g$, the polarization function $\chi^0$, and so on, are expressed as matrices. 
In this section, for the sake of simplicity, we treat them as if they were scalar quantities. One can easily extend the proof to the case where they are matrices.
$\Sigma_t  = | g^{(p)} | ^2 \chi^{t}_{\rm{DFT}}$ is rewritten as 
\begin{widetext}
\begin{eqnarray}
\Sigma_t \ &=&\  | g^{(p)}|^2 \frac{\chi^0_t} { 1 - \tilde{W}^{(p)} \chi^0_t}   \nonumber \\ 
                 &=& \  | g^{(b)}|^2   \frac{1} {1 - \tilde{v} \chi^0_r }    \frac{\chi^0_t} { 1 - \tilde{W}^{(p)} \chi^0_t}
                           \frac{1} {1 - \tilde{v} \chi^0_r }  \nonumber \\ 
                 &=& \  | g^{(b)}|^2  \left( 1 +  \frac{\tilde{v} \chi^0_r} {1 - \tilde{v} \chi^0_r } \right) 
                           \frac{\chi^0_t} { 1 - \tilde{W}^{(p)} \chi^0_t}
                           \left( 1 +  \frac{\tilde{v} \chi^0_r} {1 - \tilde{v} \chi^0_r } \right ) \nonumber \\ 
                 &=&\  | g^{(b)}|^2  \left( 1 +   \chi^0_r\tilde{W}^{(p)} \right ) 
                           \frac{\chi^0_t} { 1 - \tilde{W}^{(p)} \chi^0_t}
                           \left( 1 +   \tilde{W}^{(p)} \chi^0_r \right )  \nonumber \\ 
                 &=& \  | g^{(b)}|^2 \biggl[  \   \frac{\chi^0_t} { 1 - \tilde{W}^{(p)} \chi^0_t} 
                          +   \chi^0_r   \frac{  \tilde{W}^{(p)}   } { 1 - \tilde{W}^{(p)} \chi^0_t}   \chi^0_t 
                          +   \chi^0_t   \frac{  \tilde{W}^{(p)}  } { 1 - \tilde{W}^{(p)} \chi^0_t}  \chi^0_r  
                          +     \chi^0_r  \tilde{W}^{(p)} \frac{  \chi^0_t   } { 1 - \tilde{W}^{(p)} \chi^0_t} \tilde{W}^{(p)} \chi^0_r 
                          \ \biggr] \nonumber \\ 
                  &=&\  | g^{(b)}|^2 \biggl[  \    \chi^0_t + \chi^0_t \tilde{W}^{(f)} \chi^0_t   
                          +   \chi^0_r \tilde{W}^{(f)}   \chi^0_t  +   \chi^0_t \tilde{W}^{(f)} \chi^0_r  
                          +   \chi^0_r  \tilde{W}^{(p)} \frac{\chi^0_t} { 1 - \tilde{W}^{(p)} \chi^0_t} \tilde{W}^{(p)} \chi^0_r
                          \ \biggr].       
\end{eqnarray}
\end{widetext}
Similarly, $\Sigma_r = | g^{(b)} | ^2 \chi^{r}_{\rm{DFT}}$ is rewritten as 
\begin{eqnarray}
\Sigma_r = | g^{(b)}|^2 \frac{\chi^0_r} { 1 - \tilde{v} \chi^0_r}  
                 = | g^{(b)}|^2\biggl[  \    \chi^0_r + \chi^0_r \tilde{W}^{(p)} \chi^0_r     \ \biggr].  \nonumber \\ 
\end{eqnarray}
Using the equality 
\begin{eqnarray}
 \tilde{W}^{(p)} +   \tilde{W}^{(p)} \frac{\chi^0_t} { 1 - \tilde{W}^{(p)} \chi^0_t} \tilde{W}^{(p)} 
 =  \frac{\tilde{W}^{(p)} } { 1 - \tilde{W}^{(p)} \chi^0_t}   = \tilde{W}^{(f)},  \nonumber \\ 
 \end{eqnarray}
one can show that $\Sigma_t + \Sigma_r$ is expressed as 
\begin{eqnarray}
\Sigma_t +\Sigma_r  &=& | g^{(b)}|^2\biggl[  \    \chi^0_t + \chi^0_r  
   +  \bigl( \chi^0_t + \chi^0_r  \bigr ) \tilde{W}^{(f)} \bigl( \chi^0_t + \chi^0_r  \bigr )     \ \biggr] \nonumber \\ 
    &=&| g^{(b)}|^2\biggl[  \     \chi^0 +  \chi^0    \tilde{W}^{(f)} \chi^0   \ \biggr]  \nonumber \\ 
    &=&| g^{(b)}|^2 \chi_{\rm {DFT}}, 
\end{eqnarray}
which agrees with the expression for $\Sigma$ in Eq.~(\ref{Eq.sigma}).

\section{Smallness of electron-phonon vertex correction in downfolding procedure}
\label{app:vertex}

Due to the high phonon frequency $\sim 0.1$ eV, 
which is comparable to the typical electronic kinetic energy $\sim 0.5$ eV, the Migdal theorem~\cite{Migdal_theorem} is violated in the $\A3C60$ systems. 
Therefore, we need a careful consideration about the vertex corrections. 
In this Appendix, we argue that, as far as the processes involving the high-energy electrons are concerned,  
the electron-phonon vertex corrections are small.

\begin{figure}[tb]
\vspace{0cm}
\begin{center}
\includegraphics[width=0.32\textwidth]{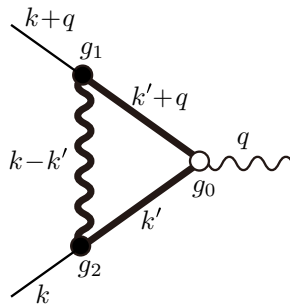}
\caption{
First-order electron-phonon vertex correction diagram.
The open (closed) circle represents the bare (dressed) electron-phonon coupling.
The bold (thin) solid and wavy lines indicate the dressed (bare) electron and phonon propagators, respectively.   
}
\label{Fig.vertex_correction}
\end{center}
\end{figure}

To see this, let us consider the ``first-order" vertex correction diagram in Fig.~\ref{Fig.vertex_correction}. 
For simplicity, we assume that the multiple intramolecular phonon modes are represented by 
 a single Einstein phonon branch with the frequency $\omega_0$
and that the electron-phonon vertex $g$ has no momentum dependence (or the electron-phonon coupling is local). 
Then, the inclusion of the diagram in Fig.~\ref{Fig.vertex_correction} gives the correction to the bare electron-phonon vertex $g_0$ 
as $g_0 \rightarrow g_0( 1 + \gamma )$ with $\gamma$ being a dimensionless quantity given by 
\begin{eqnarray}
 \gamma = - \frac{T}{N_{\bf k} }  \sum_{k'} g_1 g_2 D(k-k') G(k') G(k' \! +q), 
 \label{Eq.result_vertex_correction}
\end{eqnarray}
where $g_i$'s, $D$, and $G$ are the dressed electron-phonon vertices, phonon Green's function and electron Green's function respectively. $g_i$'s and $G$ have orbital indices, while we do not show them for simplicity. 
$k$ [$q$] represents a set of the momentum and the fermionic [bosonic] Matubara frequency $k = ({\bf k}, \nu_n)$ [$q= ( {\bf q}, \omega_n)$].
$T$ is the temperature and $N_{\bf k}$ is the number of ${\bf k}$-points. 
Note that this diagram is of first order with respect to $D$, however, it contains the higher order diagrams with respect to the bare phonon Green's function $D_0$.  In the downfolding procedure, the low-energy processes are excluded, therefore, the two electron Green's function in 
Eq.~(\ref{Eq.result_vertex_correction}) should be a combination of $G_H$ and $G_H$ or of $G_H$ and $G_L$, 
where $G_H$ ($G_L$) is the propagator of the high-energy (low-energy) electrons\cite{note_vertex}.
Then, the typical order of $\gamma$ associated with the downfolding is given by  $| \gamma | \sim 2g_1g_2 / \omega_r  \times 1/ \Delta E$ with the renormalized phonon frequency $\omega_r$ and the typical particle-hole excitation energy scale involving high-energy degrees of freedom $\Delta E$.  
Here, to derive this expression, we have employed the fact that the typical order of the convolution of $G_H$ and $G_H$ or $G_H$ and $G_L$ is $\sim 1/\Delta E$. 
In the case of the alkali-doped fullerides, $\Delta E$ is at least $\sim 1$ eV.  
$2g_1g_2 / \omega_r$ is nothing but the static part of the fully-screened phonon-mediated interaction.
If $g_1$ and $g_2$ are the coupling between phonons and low-energy electrons, it corresponds to $U^{(f)}_{\rm ph}(0)$, $U^{\prime(f)}_{\rm ph}(0)$ and $J^{(f)}_{\rm ph}(0)$ in Table.~\ref{tab_ph_int}.  
While we do not estimate the coupling between the phonon and the high-energy electrons,  
we expect the order of the phonon-mediated interactions involving the high-energy electrons is the same as that of $U^{\prime(f)}_{\rm ph}(0)$ and $J^{(f)}_{\rm ph}(0)$.
In addition, in the diagrams considered in the downfolding procedure, the orbital indices for $g_1$ and $g_2$ are usually different. 
This is because one is the coupling to electron state and the other is the coupling to the hole state, 
which would make $2g_1g_2 / \omega_r$ smaller. 
In any case, $2g_1g_2 / \omega_r$  will be at most $\sim 0.1$ eV. 
As a result, the correction $\gamma$ associated with the downfolding will take a small value $\gamma < 0.1$.

In conclusion, the neglect of the electron-phonon vertex corrections in the model-derivation step as in the case of the cDFPT is justified. 
However, we note that we still need a careful treatment for the vertex corrections in the model-analysis step since 
the vertex corrections in the $t$-subspace is not negligible any more.  

\bibliographystyle{apsrev}
\bibliography{condmat}

\begin{thebibliography}{93}
\bibitem[{not({\natexlab{b}})}]{note2}
\bibinfo{note}{Strictly speaking, this expression [Eq.~(\ref{Eq:deln})] is
  valid only when the ionic potential $V_{\rm{ion}}$ is purely local. In
  practice, we employ the pseudopotential, which has non-local part. In this
  case, one has to introduce three-point response functions in the formulation,
  however, it does not change the outline presented in this section.}
\bibitem[{not({\natexlab{c}})}]{note_Ag}
\bibinfo{note}{The two $A_g$ modes can couple to the $t_{1u}$ electrons,
  however, they do not exist in this range; the experimentally observed
  frequencies are 496 and 1470 cm$^{-1}$ [\onlinecite{Bethune1991181}].}
\bibitem[{not({\natexlab{d}})}]{note_ph_sum}
\bibinfo{note}{In principle, the sum over $\nu$ in
  Eq.~(\ref{Eq.Vpartial_result_sec}) runs from 1 to 189. However, we omit the
  contribution from the lowest 9 branches ($\nu \! =\! 1$-9). They correspond
  to the acoustic modes, the librations, and the alkali-ion vibrations at the
  octahedral sites.~\cite{PhysRevB.88.054510} Within the present calculation,
  some of these vibrations have imaginary frequencies. The neglect can be
  justified since the couplings between these modes and the $t_{1u}$ electrons
  are small [\onlinecite{RevModPhys.69.575,Gun_book}].}
\bibitem[{not({\natexlab{e}})}]{note_vertex}
\bibinfo{note}{If the off-diagonal Green's function $G_{HL}$ is nonzero, we
  also have to take into account the processes involving $G_{HL}$. However, the
  off-diagonal Green's functions are usually very small compared to the
  diagonal ones, thus the effects of $G_{HL}$ on the vertex correction $\gamma$
  are small.}
\end{thebibliography}

\end{document}